\documentclass[amsfonts, amssymb, amsmath, twocolumn, showkeys, nofootinbib, reprint,floatfix, prx, superscriptaddress]{revtex4-2}

\usepackage[english]{babel}
\usepackage[colorinlistoftodos, color=green!40,prependcaption]{todonotes}
\usepackage{graphicx}
\usepackage{dcolumn}
\usepackage{bm}
\usepackage{float}
\usepackage{gensymb}
\usepackage{xcolor}
\usepackage{amsmath}
\usepackage{amssymb}
\usepackage{units}
\usepackage{placeins}
\usepackage{ulem}
\usepackage{hyperref}
\usepackage{bbm}
\usepackage{multirow}
\usepackage{amsfonts}
\usepackage{amsmath}
\usepackage{amssymb}
\usepackage{verbatim}
\usepackage{cat}

\usepackage{lipsum}
\def\eia~{EuIn$_2$As$_2$}
\usepackage{comment}

\def\el{\eta_\parallel}
\def\et{\eta_\perp}

\begin{document}

\preprint{APS/123-QED}

\title{Goldstone mode of the broken helix in $\text{U}(1)$ magnet \eia~}

\author{A. Liebman-Pel\'{a}ez}
\email{liebmana@berkeley.edu}
\affiliation {Department of Physics, University of California, Berkeley, California 94720, USA}
\affiliation {Materials Science Division, Lawrence Berkeley National Laboratory, Berkeley, California 94720, USA}
\author{S. J. Garratt}
\affiliation {Department of Physics, University of California, Berkeley, California 94720, USA}
\author{V. Sunko}
\author{Y. Sun}
\affiliation {Department of Physics, University of California, Berkeley, California 94720, USA}
\affiliation {Materials Science Division, Lawrence Berkeley National Laboratory, Berkeley, California 94720, USA}
\author{J. R. Soh}
\affiliation {Quantum Innovation Centre, Agency for Science Technology and Research, 2 Fusionopolis Way, Singapore, 138634, Singapore}
\author{D. Prabhakaran}
\author{A. T. Boothroyd}
\affiliation {Department of Physics, University of Oxford, Clarendon Laboratory, Oxford, OX1 3PU, UK}
\author{J. Orenstein}
\affiliation {Department of Physics, University of California, Berkeley, California 94720, USA}
\affiliation {Materials Science Division, Lawrence Berkeley National Laboratory, Berkeley, California 94720, USA}

\begin{abstract}
Goldstone modes acquire a frequency gap in the presence of perturbations that break the underlying continuous symmetry. Here, we study the response of a spin-based Goldstone mode to strain and magnetic field in the broken helix, a multi-$\bm{Q}$ phase of \eia~. Optical polarimetry with spatial and temporal resolution allows us to access information about both the structure and frequency of optically excited spin-wave modes under different strain conditions. We observe nearly uniform spin precession characteristic of a Goldstone mode only when magnetic field dominates over strain. In this regime, the frequency depends linearly on the applied field. A symmetry analysis for predicting the mode frequency near zero field demonstrates that the observed scaling is of the lowest allowed order. This work thus demonstrates the connections between magnetic symmetries and the frequency dependence of the Goldstone mode in an external field, and illustrates the power of our technique for studying the dynamics of complex magnets.
\end{abstract}

\maketitle

\begin{figure*}
    \centering
    \includegraphics[scale=1]{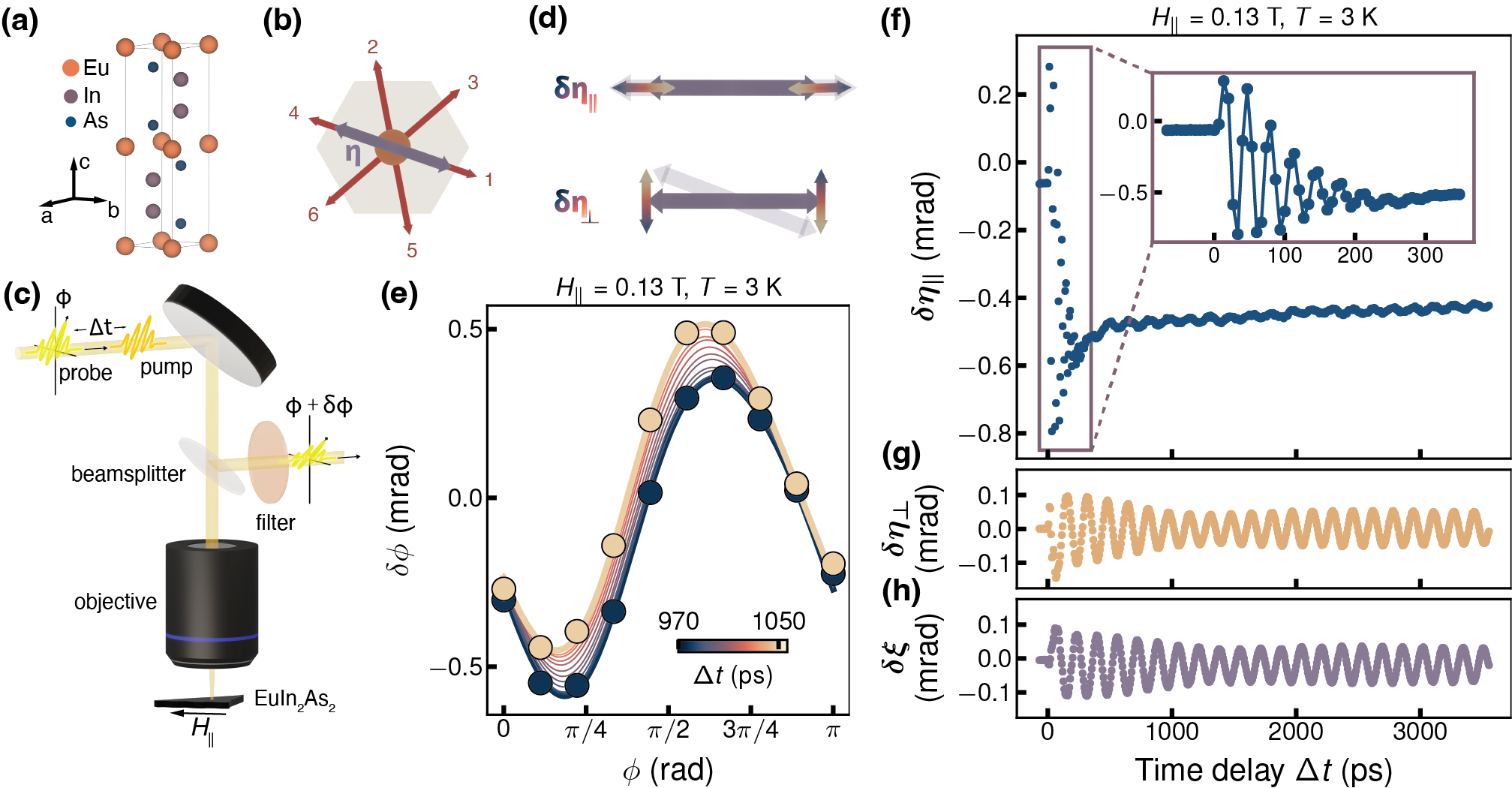}
    \caption{\textbf{(a)} Structural unit cell of EuIn$_2$As$_2$. \textbf{(b)} Top view of the broken helix, with orange arrows representing the spin orientation in each Eu plane of the magnetic unit cell, labeled 1-6. Purple double headed arrow represents the orientation of the nematic director $\bm{\eta}$. \textbf{(c)} Schematic of the optical setup used for measurements of the polarization rotation $\delta\phi$ as a function of incident polarization $\phi$, time delay between pump and probe $\Delta t$, and an in-plane magnetic field $H_\parallel$. The sample is held at $T = \unit[3]{K}$ and both beam spots are $\approx\unit[5]{\mu m}$ in diameter. \textbf{(d)} Relation between the equilibrium nematic director (solid purple) and the time dependent fluctuations $\delta\el$ and $\delta\et$ \textbf{(e)} Polarization rotation as a function of incident polarization for different values of the time delay at $H_\parallel\approx\unit[0.13]{T}$. Circles are data points and solid lines are best fits. Coherent pump-induced spin-wave dynamics at $H_\parallel\approx\unit[0.13]{T}$ decomposed into \textbf{(f)} longitudinal nematic $\delta\el$, \textbf{(g)} transverse nematic $\delta\et$, and \textbf{(h)} MOKE $\delta\gamma$ channels plotted on the same scale; inset on \textbf{(f)} shows high frequency oscillations in $\delta\el$ at small time delays.}
    \label{fig:setup_corot_timedep}
\end{figure*}

Systems that spontaneously break a continuous symmetry give rise to a Goldstone mode \cite{kardar_statistical_2007}. In the context of spin systems that possess $\text{U}(1)$ symmetry, i.e. their Hamiltonian is invariant under rotations within a plane in spin space, the Goldstone mode is a smooth rotation of the spin configuration with a vanishing frequency at long wavelengths. These properties are universal to all magnetic ground states that spontaneously break $\text{U(1)}$ \cite{halperin_hydrodynamic_1969, halperin_hydrodynamic_1977, fishman_goldstone_1996}. However, in physical systems $\text{U}(1)$ symmetry is rarely perfectly preserved, and the Goldstone mode acquires a frequency gap $f_G \neq 0$ \cite{macneill_gigahertz_2019, bedoya-pinto_intrinsic_2021, tymoshenko_pseudo-goldstone_2017, grassi_higgs_2022}. This Letter addresses, experimentally and theoretically, how the symmetry of the spin configuration determines the effect of $\text{U}(1)$ breaking perturbations on the Goldstone mode structure and frequency.

One way to break $\text{U}(1)$ is by applying a magnetic field in the spin plane ($\bm{H}_{\parallel}$). The $\bm{H}_{\parallel}$-dependence of $f_G$ has been studied theoretically within Heisenberg exchange models that yield spin structures with a single modulation vector $\bm{Q}$ \cite{cooper_theory_1962, cooper_spin-wave_1963, zaliznyak_modification_1995, andreev_symmetry_1980}. It was shown that $f_G \sim H_{\parallel}^{N/2}$, where $N=2\pi/Q$ is the number of spins in the magnetic unit cell. However, much less is understood about the response of multi-$\bm{Q}$ spin structures \cite{ziman_linear_1986, dos_santos_modeling_2020, diep_quantum_2022}. To our knowledge, whether there exists a simple, universal relationship between the scaling of $f_G$ with $\bm{H}_\parallel$ and magnetic structure is not known. While previous work has addressed this issue in systems with a net magnetic moment \cite{kittel_theory_1948, bauer_soft_2023}, a unified picture for systems with no net moment is lacking. An additional way of breaking $\text{U}(1)$ symmetry is through uniaxial strain, which is often unavoidable because it is generated during the crystal growth process.

Here, we utilize time-resolved optical polarimetry to identify the modifications to the Goldstone mode under both types of symmetry breaking fields, $\bm{H}_{\parallel}$ and strain. For our study we chose \eia~, a material which possesses a near-perfect $\text{U}(1)$ symmetry \cite{donoway_multimodal_2024}. Its ground state, termed the `broken helix' \cite{riberolles_magnetic_2021, soh_understanding_2023, gen_incommensurate_2024}, is characterized by interpenetrating antiferromagnetic and six-fold helical modulations, making it one of the simplest multi-$\bm{Q}$ structures, with ordering at two parallel wavevectors. While the spin dynamics of \eia~ have been studied at zero field \cite{liu_exploring_2024} and in applied out-of-plane field \cite{wu_spin_2023}, there have been no reports on the influence of in-plane field prior to this work.

We first study the interplay of strain and magnetic field in determining the structure of the lowest-frequency mode. We find that uniform spin precession is approximately preserved when $\bm{H}_{\parallel}$ is weak compared to the exchange interaction, but strong compared with the strain. Next, we investigate the dependence of $f_G$ on larger magnetic fields. Within a Heisenberg exchange model, the Goldstone mode frequencies of Ne\'el antiferromagnetic and six-fold helical structures are expected to vary as $H_{\parallel}$ and $H_{\parallel}^3$, respectively; it is \textit{a priori} unclear if the broken helix, which has both modulations, inherits either of those behaviors. Strikingly, we find that $f_G$ depends linearly on $H_{\parallel}$, with little deviation up to the saturation field - evidently the broken helix behaves like an antiferromagnet rather than a helix. Motivated by this result, we derive the general symmetry-imposed constraints on the field dependence of $f_G$, and apply them to three representative structures -- antiferromagnet, helix, and broken helix -- clarifying the origin of their respective field dependencies. Together, our findings identify the universal and non-universal modifications of Goldstone modes caused by symmetry-breaking fields.

The structure of \eia~ is shown in Fig.~\ref{fig:setup_corot_timedep}a. The localized moments of Eu$^{2+}$ ions ($S=7/2$, $L=0$) provide magnetic degrees of freedom, which order in a sequence of transitions. Below $T_{N1} \approx \unit[17.5]{K}$, the spins align ferromagnetically within each Eu plane and develop a modulation at $\bm{Q}_1 = (0,0,1/3)$. Since \eia~ has two Eu per structural unit cell, this corresponds to order with six Eu per magnetic unit cell. Below $T_{N2}\approx\unit[16]{K}$, an additional, interpenetrating antiferromagnetic modulation develops at $\bm{Q}_2 = (0, 0, 1)$, yielding the broken helix ground state (Fig.~\ref{fig:setup_corot_timedep}b). The broken helix breaks $C_{3z}$ symmetry, giving rise to a nematic order parameter $\bm{\eta}$ (purple double-headed arrow in Fig.~\ref{fig:setup_corot_timedep}b).  Its two components can be expressed in terms of an amplitude and angle, $\bm{\eta} = \eta\left(\cos2\theta, \sin2\theta\right)$, and both can be observed through the anisotropy in reflectivity induced by $\bm{\eta}$ ~\cite{donoway_multimodal_2024}.

\begin{figure*}
    \centering
    \includegraphics[scale=1]{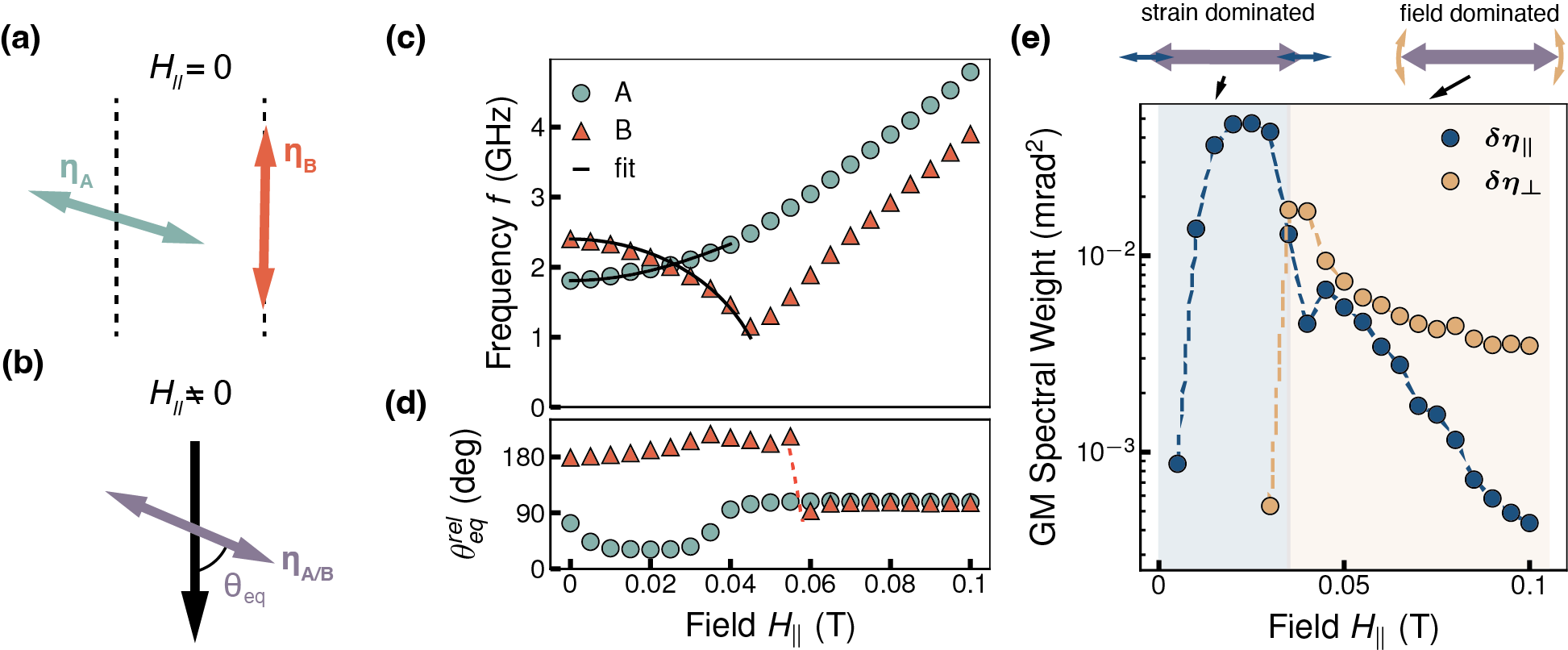}
    \caption{\textbf{(a)} Depiction of orientation of $\bm{\eta}_\text{eq}$ at positions A and B in zero field, relative to the axis of the applied field. \textbf{(b)} Definition of $\theta_\text{eq}$, angle between $\bm{H}_\parallel$ (black arrow) and $\bm{\eta}_\text{eq}^A$ or $\bm{\eta}_\text{eq}^B$. \textbf{(c)} Field dependence of the Goldstone mode frequency at positions A and B in the low-field regime at $T=\unit[3]{K}$. Black lines indicate fits to $f(H) = \sqrt{f^2(0) + AH^2}$, where $A$ is the only fitting parameter. \textbf{(d)} $\theta_\text{eq}$ as a function of applied field $H_\parallel$; $\theta_\text{eq}^{(A)}(0) \approx 73$ degrees, $\theta_\text{eq}^{(B)}(0) \approx 179$ degrees. \textbf{(e)} Integrated weight of the Goldstone mode in $\delta\el$ and $\delta\et$ channels at position A. At low fields, strain dominates and the mode has primarily $\delta\el$ spectral weight. Past the spin-flop transition, the field dominates and the mode exhibits stronger $\delta\et$ spectral weight, indicating a near-uniform precession.}
    \label{fig:strain}
\end{figure*}

To measure spin dynamics in the broken helix, we first optically excite spin-waves with an ultrafast pump pulse (\unit[620]{nm}; \unit[510]{$\mu$J/cm$^2$}). Spin precession induces time-dependent changes to $\bm{\eta}$, such that $\bm{\eta}(\Delta t) = \bm{\eta}_\text{eq} + \delta\bm{\eta}(\Delta t)$, which modulate the anisotropic reflectivity. We detect $\delta\bm{\eta}$ by measuring the polarization rotation of a probe pulse (\unit[580]{nm}; \unit[2120]{$\mu$J/cm$^2$}) reflected at normal incidence, which arrives at the sample a time $\Delta t$ after the pump (Fig.~\ref{fig:setup_corot_timedep}c). Both pump and probe beams are focused to an approximately $\unit[5]{\mu m}$ diameter spot on the sample. The time-dependent changes of the amplitude and orientation of $\bm{\eta}$ correspond to the components of $\delta\bm{\eta}(\Delta t)$ that are longitudinal ($\delta\el$) and transverse ($\delta\et$) with respect to the equilibrium nematic director $\bm{\eta}_\text{eq}$, respectively (Fig.~\ref{fig:setup_corot_timedep}d). A uniform spin precession yields oscillations of $\delta\et$, while the relative motion of spins with respect to one another results in oscillations of $\delta\el$. It is therefore beneficial to decompose  $\delta\bm{\eta}(\Delta t)$ as\footnote{We stress that $\eta \neq \eta_\text{eq} + \delta\eta$ and $\theta \neq \theta_\text{eq} + \delta\theta$.}:
\begin{equation}
\delta\bm{\eta}(\Delta t) = \delta\eta (\Delta t) \left (\begin{matrix} \cos2(\delta\theta (\Delta t) - \theta_\text{eq}) \\ \sin2(\delta\theta(\Delta t) - \theta_\text{eq})\end{matrix}\right) \equiv \left (\begin{matrix} \delta\el (\Delta t) \\ \delta\et (\Delta t)\end{matrix}\right ),
\label{eq:nematic}
\end{equation}
where $\theta_\text{eq}$ is the equilibrium angle. In addition to changes of $\bm{\eta}$, the excited broken helix acquires a time-dependent magnetization, similar to antiferromagnets \cite{rezende_introduction_2019, sun_dipolar_2024}. The component of the dynamic magnetization along $\bm{c}$ ($M_c$) contributes to the polarization rotation through the polar magneto-optical Kerr effect (MOKE), which we parametrize by $\delta\xi(\Delta t)$. 

We are able to simultaneously measure and distinguish $\delta\bm{\eta}$ and $\delta\xi$ because they contribute to the symmetric and antisymmetric parts of the reflectivity tensor, respectively. For a probe pulse with initial polarization angle $\phi$, the rotation $\delta\phi$ detected by the balanced optical bridge method is given to first order by \cite{SI},
\begin{equation}
    \label{eq:corotations}
    \delta\phi(\phi, \Delta t) = \delta\eta(\Delta t)\sin(2(\phi - \delta\theta(\Delta t))) + \delta\xi(\Delta t).
\end{equation}
From the sinusoidal variations with $\phi$ we can access the dynamics of the nematic order parameter, while via $\delta\xi$ we can access changes in $M_c$. An example of the data thus measured at $T=\unit[3]{K}$ is shown in Fig.~\ref{fig:setup_corot_timedep}e, where we plot $\delta\phi(\phi, \Delta t)$ for various values of $\Delta t$ at fixed magnetic field. It is clear that Eq.~\eqref{eq:corotations} is obeyed, and that both the sinusoidal and $\phi$-independent components depend on $\Delta t$. By fitting such data to Eq.~\eqref{eq:corotations} we extract the time dependence of the three parameters $\delta\el$,  $\delta\et$ and $\delta\xi$ (\textit{c.f.}   Eq.~\eqref{eq:nematic}). Access to all three yields insight into both the structure and frequency of spin-wave modes.

In Fig.~\ref{fig:setup_corot_timedep}f-h we show the time-dependence of $\delta\el$, $\delta\et$, and $\delta\xi$ extracted from measurements such as those in Fig.~\ref{fig:setup_corot_timedep}e. The most prominent feature in the first $\unit[300]{ps}$ is a highly damped oscillation at $\sim\unit[30]{GHz}$ (Fig.~\ref{fig:setup_corot_timedep}f, and the inset). Because this mode appears only in the $\delta\el$ channel, we associate it with the relative motion of spins within a magnetic unit cell, and identify it as an optical spin-wave mode. In addition, we observe a second, long-lived $\sim \unit[6]{GHz}$ oscillation in all channels. The amplitude at this frequency is much stronger in the $\delta\et$ channel than in the $\delta\el$ channel, indicating that this mode is characterized by a near-uniform precession of spins, as expected for a Goldstone mode. We note that the envelopes of $\delta\et$ and $\delta\xi$ reveal a slight beating pattern; we discuss possible origins in the Supplemental Materials (SM) ~\cite{SI}.

Having established the methodology to analyze the structure and frequency of spin-waves, we now investigate the dependence of the low frequency mode on strain and magnetic field. \eia~ is the ideal candidate for this study: as-grown crystals possess an in-built strain, whose influence on $\bm{\eta}_\text{eq}$ has been extensively characterized in the same crystal of \eia~ that we use for this study. In Ref.~\cite{donoway_multimodal_2024} it was shown that strain varies smoothly across the crystal, and pins the local orientation of $\bm{\eta}_\text{eq}$. The spatial resolution of our optical probe allows us to access a continuum of strain orientations in a single sample by scanning the position of the laser focus. We note that this would be impossible in experiments with deliberately applied strain: a new sample would be needed for each strain orientation.

By scanning the laser focus, we measure the equilibrium nematic orientation and the low-field spin-wave dynamics as a function of strain direction \cite{SI}. Here, we concentrate on two representative positions, A and B, chosen so that the field is applied approximately parallel to $\bm{\eta}_\textnormal{eq}$ at B, and close to perpendicular at A (Fig.~\ref{fig:strain}a,b). At each position, the frequency of the Goldstone mode in the field range $H_\parallel = \unit[0-0.1]{T}$ was extracted from simultaneous fits to the Fourier transforms of $\delta\el$, $\delta\et$, and $\delta\xi$  (see the SM~\cite{SI} for technical details). In Fig.~\ref{fig:strain}c we show the frequency vs. $H_\parallel$ for positions A and B, and in Fig.~\ref{fig:strain}d the field evolution of $\theta_\text{eq}^{(A,B)}$, the equilibrium nematic orientation at each position. At both positions, $\bm{\eta}_\text{eq}$ orients approximately perpendicular to the field above a spin-flop field, $H_\text{f}$, that is position dependent ($H_\text{f}(A)=\unit[0.04]{T}$, $H_\text{f}(B)=\unit[0.06]{T}$). The spin-flop transition corresponds to the broken helix unpinning from the strain-induced easy-axis. The frequencies at both positions are linear in $H_\parallel$ for $H_\parallel\gtrsim H_\text{f}$. At lower fields, the dependence of frequency on field is strongly modified by strain -- increasing with field at position A and decreasing at position B. Thus the magnetic field increases the effective anisotropy at position A and decreases it at position B, consistent with the orientations depicted in Fig.~\ref{fig:strain}a and the observation that $\bm{\eta}_{eq}$ orients perpendicular to $H_{\parallel}$ at high fields. These features, as well as the fact that the flop transition is both sharper and at a higher field for the field applied closer to the easy-axis, are reproduced by basic models of antiferromagnetic and helical magnets \cite{SI}.

Our ability to measure both the longitudinal and transverse oscillations of the nematic director allows us to characterize the change in mode structure through the crossover from the strain-dominated ($H_\parallel<H_\text{f}$) to the field-dominated regime ($H_\parallel>H_\text{f}$). In Fig.~\ref{fig:strain}e we plot the integrated spectral weight of the Goldstone mode at position A for the $\delta\el$ and $\delta\et$ channels. At low fields, $\delta\el$ is orders of magnitude larger than $\delta\et$, indicating that intra-unit cell motion of spins is dominant, changing the nematic amplitude. The picture reverses as the broken helix unpins from the strain for $H_\parallel>H_\text{f}$, where the rapid increase in spectral weight of $\delta\et$ relative to $\delta\el$ indicates that the mode is characterized by a near-uniform rotation of the nematic order. The lowest frequency mode with broken $\text{U}(1)$ symmetry is therefore only similar to an idealized Goldstone mode when the dominant symmetry-breaking parameter is $H_{\parallel}$, but not when strain dominates.

\begin{figure}
    \centering
    \includegraphics[scale=1]{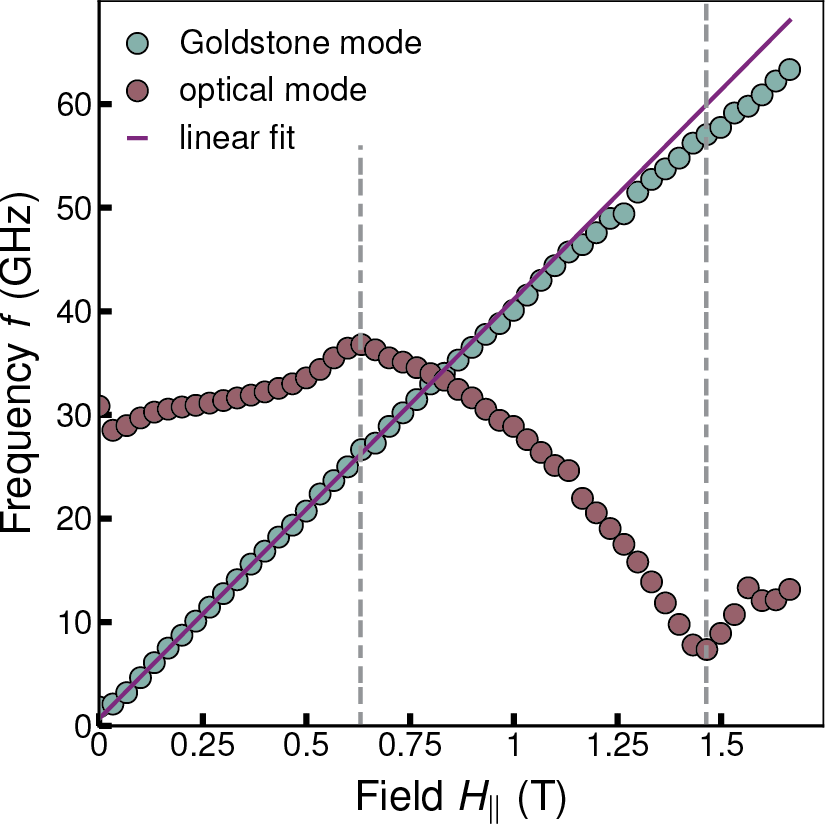}
    \caption{Magnetic field dependence of the frequency of the optical and Goldstone modes of the broken helix at $T=\unit[3]{K}$. Vertical dashed lines mark $H_\text{hf}$ and $H_\text{sat}$. The purple line shows a linear fit to the Goldstone mode from $H = 0 \rightarrow H_\text{hf}$.}
    \label{fig:field_dependence}
\end{figure}

Having demonstrated the applicability of the Goldstone mode picture in the field-dominated regime, we now expand the study of mode frequencies to higher fields. In Fig.~\ref{fig:field_dependence} we show the field evolution of the optical and Goldstone modes at position A over $H_\parallel = \unit[0-1.75]{T}$, which extends well beyond the saturation field, $H_\text{sat}\approx \unit[1.46]{T}$. The frequency of the optical mode increases gradually up to $H_\text{hf} \approx \unit[0.63]{T}$, and then decreases to a minimum at the saturation field $H_\text{sat}$. The peak at $H_\text{hf}$ coincides with the onset of rapid changes in $\theta_\text{eq}$ \cite{SI}, suggesting that it corresponds to the helix-to-fan transition observed in magnetization measurements on \eia~ \cite{soh_understanding_2023}. Most striking is the observation that the frequency of the Goldstone mode remains linear in $H_\parallel$ even as the field exceeds $H_\text{hf}$, where the spin structure undergoes a massive reorganization. 

Despite the complexity of the broken helix under the influence of applied in-plane fields, the field dependence of the Goldstone mode is remarkably simple, mirroring that of an isotropic antiferromagnet. We now present a symmetry-based argument to explain this surprising observation. Low-energy semiclassical dynamics at zero wavevector can be described in terms of the amplitudes $m^1_j$ and $m^2_j$ of spin fluctuations along directions orthogonal to the ground state spin orientations. Here $j=1,\ldots,N$ labels sites within the magnetic unit cell. The  zero-wavevector linearized dynamics are governed by,
\begin{align}
    \frac{d}{dt} m^a_j = 2\pi i\sum_{b=1}^2\sum_{j=1}^N D^{ab}_{jk} m^b_k, \label{eq:linearizeddynamics}
\end{align}
where $D^{ab}_{jk}$ are the components of a $2N \times 2N$ dynamical matrix $D$. The mode frequencies are the eigenvalues of $D$, and these eigenvalues come in oppositely signed pairs $\pm f_{\nu}$. In a magnetic field $\bm{H}$ the ground state spin orientations are modified and hence $D$ depends on $\bm{H}$. 

Our aim is to determine how the mode frequencies $f_{\nu}(\bm{H})$ are constrained by the symmetries of the zero-field ground state configuration. The essential argument, which we outline here and expand on in the SM~\cite{SI}, is based on the fact that mode frequencies are unchanged when applying $\bm{H}$ along symmetry-related directions.  If the spin configuration distorts smoothly on increasing field, it can be expanded in powers of $\bm{H}$ about the zero-field configuration. The same holds for the dynamical matrix, and its characteristic polynomial $p(f,\bm{H})=\text{det}(f I-D(\bm{H}))$, whose roots are the mode frequencies ($I$ is the identity matrix). The mode frequencies, on the other hand, need not be analytic functions of the field. Because eigenvalues come in oppositely signed pairs, we have
\begin{align}
    p(f,\bm{H}) = \prod_{\nu=1}^N \big( f^2 - f^2_{\nu}(\bm{H})\big). \label{eq:cpvals}
\end{align}

We now focus on a single non-degenerate mode $\mu$; we evaluate $p(f,\bm{H})$ at $f = f_\mu(0)$, and investigate its low-field behavior. Note that the $\mu^{\textnormal{th}}$ term in the product determines the lowest order dependence of $p(f,\bm{H})$ on $\bm{H}$, because all other terms contain $\bm{H}$-independent constants $f^2_{\mu}(0) - f^2_{\nu}(0)$. Since the characteristic polynomial can be Taylor expanded in powers of $H$, we may write
\begin{equation}  
    f^2_\mu(\bm{H}) - f_\mu^2(0) \sim \sum_{uv} c_{uv,\mu}H^{u+v} (\cos\varphi)^u (\sin\varphi)^v,
    \label{eq:cpexpansion}
\end{equation}
where we have assumed that the field is in-plane, $\bm{H} = H(\cos\varphi, \sin\varphi,0)$, and we have introduced $\mu$-dependent coefficients $c_{uv,\mu}$. Crucially, the right hand side is constrained by combinations of $u$ and $v$ that respect the symmetry of the zero-field ground state configuration.

In the absence of strain we have $f_G(0)=0$ for a Goldstone mode. For the $C_{2z}$-symmetric broken helix, the characteristic polynomial has to be invariant under the change of $\varphi$ by $\pi$, which is accompanied by an additional factor of $(-1)^{u+v}$ on the right-hand side of Eq.~\eqref{eq:cpexpansion}. Clearly, only even values of $u+v$ keep $p(f, \bm{H})$ invariant, so $f^2_G(\bm{H}) \sim H^2$ is the lowest order symmetry-allowed contribution to the Goldstone mode frequency in $C_{2z}$-symmetric magnets. This accounts for the experimental observation of $f_G(\bm{H}) \sim H$ in the regime where the field dominates. It also clarifies why the broken helix and the Ne\'el antiferromagnet exhibit the same field-dependence: the rotational symmetry of the ground state is the same. To complement our symmetry analysis, in the SM~\cite{SI} we determine the conditions under which the linear contribution to $f_G(\bm{H})$ is nonzero within a microscopic spin model, confirming that it is generically nonzero for the broken helix. Eq.~\eqref{eq:cpexpansion} also accounts for the field dependence in the strain-dominated regime, when $f_\mu(0)\neq0$, and $f_\mu(\bm{H}) \sim \sqrt{f_\mu^2(0) + AH^2}$. In Fig.\ \ref{fig:strain}c we show fits of the Goldstone mode frequency to this functional form.

The behavior identified above for the broken helix is markedly different to that of a conventional helix, where for an $N$-spin magnetic unit cell one finds $f_G(\bm{H}) \sim H^{N/2}$ \cite{zaliznyak_modification_1995}. This is in spite of the fact that a contribution $f_G(\bm{H}) \sim H$ can arise from a rotationally symmetric $\bm{H}\cdot\bm{H}$ term in the right-hand side of Eq.~\eqref{eq:cpexpansion}. The scaling $f_G(\bm{H}) \sim H^{N/2}$, which we confirm numerically across a wide variety of microscopic models \cite{SI}, corresponds to the lowest-order contribution to Eq.~\eqref{eq:cpexpansion} that is allowed by the discrete symmetry of the ground state, but which does not have a continuous symmetry (e.g. $ p(f,\bm{H})\sim H^N\cos(N\varphi)$), indicating that terms with continuous symmetry vanish in $f_G(\bm{H})$. Whether this is true for \textit{all} magnetic structures is a topic for future theoretical study motivated by our work. However, the question is not meaningful for $C_{2z}$-symmetric magnets, such as the broken helix, since the $f_G(\bm{H}) \sim H$ term is the lowest one allowed by both the discrete and the continuous symmetry.

In this work, we have observed the Goldstone mode of the broken helix in \eia~ via time-resolved optical polarimetry, and studied how strain and magnetic field alter its properties. We found that near-uniform spin precession is preserved when magnetic field is the dominant symmetry-breaking parameter, but not when strain dominates. The spin-wave frequency scales linearly with an applied magnetic field, as it does in antiferromagnets, despite a considerably more complex magnetic structure. We have explained this via a general symmetry argument: an antiferromagnet and the broken helix have the same rotational symmetry. 
We emphasize that our optical technique is uniquely well-suited for studying the asymptotic behavior of the Goldstone mode frequency as $H\rightarrow 0$ for several reasons: (1) it enables access to low energies with high resolution, (2) it has a spatial resolution of $\sim \unit[5]{\mu m}$, (3) it provides information regarding the mode structure, as well as its frequency. These features can be leveraged to study the low-energy dynamics of complex magnetic systems and devices more broadly.

\vskip 0.1in

We would like to thank Ehud Altman and John Chalker for helpful discussions. This research was primarily funded by the Quantum Materials (KC2202) program under the U.S. Department of Energy, Office of Science, Office of Basic Energy Sciences, Materials Sciences and Engineering Division under Contract No. DE-AC02-05CH11231, which supported the experimental and theoretical work at the LBNL and UC Berkeley. D. P. and A. T. B. would like to acknowledge the Engineering and Physical Sciences Research Council, UK and the Oxford- ShanghaiTech collaboration project for financial support. J. O. received support from the Gordon and Betty Moore Foundation’s EPiQS Initiative through Grant No. GBMF4537 to J. O. at UC Berkeley. V. S. is supported by the Miller Institute for Basic Research in Science, UC Berkeley. S. J. G. was supported by the Gordon and Betty Moore Foundation.

\bibliographystyle{apsrev4-1}


\bibliography{EuIn2As2_magnons_references}

\counterwithout{equation}{section}
\renewcommand\theequation{S\arabic{equation}}
\renewcommand\thefigure{S\arabic{figure}}
\renewcommand\thetable{S\arabic{table}}
\renewcommand\thesection{S\arabic{section}}
\renewcommand\bibnumfmt[1]{[S#1]}
\setcounter{equation}{0}
\setcounter{figure}{0}
\setcounter{enumiv}{0}

\clearpage
\newpage

\section*{Supplemental Materials}

\section{Optical setup and experiments}\label{sec:SI_optical}

\begin{figure}[t!]
    \centering
    \includegraphics[scale=0.9]{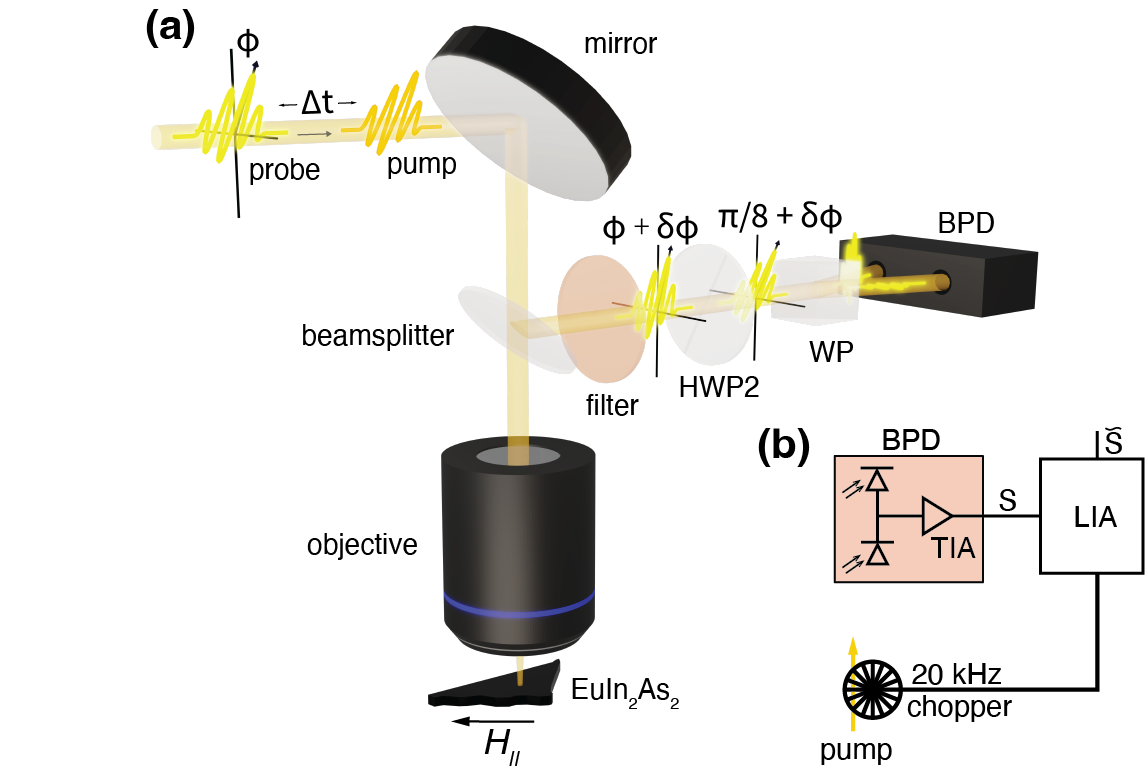}
    \caption{\textbf{(a)} Expanded schematic of the optical setup. HWP: half wave plate; WP: Wollaston prism; BPD: balanced photodiodes. HWP1, which is used to set initial polarization, is not shown. \textbf{(b)} Schematic of detection circuit; TIA: transimpedance amplifier, LIA: lock-in amplifier.}
    \label{fig:SI_optical} 
\end{figure}

An expanded schematic of the optical setup is shown in Fig.~\ref{fig:SI_optical}a. Both pump and probe pulses (pulse width $\unit[250]{fs}$; pulse rate $\unit[300]{kHz})$ have incident polarization $\phi$, set by a polarizer and half-wave plate (HWP1) (not shown in the diagram); $\phi=0^\circ$ corresponds to $p$-polarized light at the sample surface. Both beams are focused down to a diameter of $\approx \unit[5]{\mu m}$ on the sample through a 10x objective. After reflecting off the sample, the pump beam is rejected by a color filter and the probe polariztion is rotated by an angle of $\pi/8 - \phi$ by a second half-wave plate (HWP2). If the polarization state were not altered upon reflection, the polarization of the probe beam after the HWP would be an equal superposition of horizontal (H) and vertical (V) linearly polarized light. The beam is then sent through a Wollaston prism (WP), which spatially separates H and V components of the probe. Both components are focused onto separate, unbiased photodiodes of a balanced optical bridge detector. 

The photovoltage, amplified by a high-frequency transimpedance amplifier (TIA, gain:$10^5$), is thus proportional to 

\begin{equation}
    S = |E_V|^2 - |E_H|^2,
    \label{eq:SI_sig}
\end{equation}

\noindent
where $E_V$ and $E_H$ are the V and H components of the electric field at each photodiode. When H and V components have equal intensity, as would be the case for a probe beam whose polarization is unaffected by the sample, $S=0$. In contrast, a nonzero value of $S$ indicates a rotation of the polarization upon reflection. As shown in Fig.~\ref{fig:SI_optical}b, the pump intensity is modulated at $f_\text{ref} = \unit[20]{kHz}$ using an optical chopper, and $S$ is detected using a lockin-in amplifier locked to $f_\text{ref}$. Although changes to the final probe polarization can be introduced by the birefringence or ellipticity of the setup, the modulation ensures that the measurement is only sensitive to changes induced by the pump and hence drastically reduces the setup contribution, as explained below. In addition, operating at high frequencies where $1/f$ noise is small means that our measurement is at the shot noise limit. We refer to the final, modulated signal as $\tilde{S}$.

Optical anisotropies of the sample, either at equilibrium or induced by the pump, cause a rotation of the polarization of the probe upon reflection. The signal $S$ can be modeled using the Jones calculus formalism, in which the polarization state of the light is represented by a vector in the (H, V) basis and the action of each optical component is represented by a $2\times 2$ matrix. The Jones matrix for a HWP with its fast axis rotated by $\varphi$ with respect to the vertical axis is given by

\begin{equation}
    J_\text{HWP}(\varphi) = \left(\begin{matrix}-\cos2\varphi & \sin2\varphi \\ \sin2\varphi & \cos2\varphi\end{matrix}\right).
    \label{eq:SI_jones_HWP}
\end{equation}

\noindent
If the sample were optically isotropic, its Jones matrix would be proportional to the identity. However, in the presence of nematic order and/or a magnetization along the optic axis, the system becomes optically anisotropic. The nematic order leads to birefringence (difference in reflectivity between H and V polarizations). On the other hand, the magnetization leads to a polar magneto-optical Kerr effect (difference in reflectivity between left-~ and right-circular polarizations). We may then represent the the sample in the (H, V) basis as

\begin{equation}
    J_\text{samp}(r, \eta, \theta, \xi) = R(-\theta)\left(\begin{matrix}r + \eta & \xi \\ -\xi & r - \eta\end{matrix}\right)R(\theta),
    \label{eq:SI_jones_sample}
\end{equation}

\noindent
where $r$ is the reflectivity of the sample, $\eta$ is the birefringence, $\theta$ is the angle of the nematic director relative to the lab coordinates, $R(\theta) = \left(\begin{matrix}\cos\theta & \sin\theta \\ -\sin\theta & \cos\theta\end{matrix}\right)$ represents a clockwise rotation, and $\xi$ is the magneto-optical Kerr effect. It is in this sense that we express the birefringence as a nematic order parameter $\bm{\eta} = \eta\left(\begin{matrix}\cos2\theta \\ \sin2\theta \end{matrix}\right) \equiv \left(\begin{matrix}\eta_x \\ \eta_y\end{matrix}\right)$ in the main text. We have chosen to set the lab coordinates $(x, y)$ such that $p$-polarized light lies along $x$, i.e.~ $(x,y)$ corresponds to $(V, H)$. This was to ensure that $\bm{H}_\parallel \parallel \hat{x}$. For a magnetic structure $\bm{S}_i = (S_i^x, S_i^y, S_i^z)$, $\bm{\eta}$ can be related directly to the magnetic structure,
\begin{equation}
\bm{\eta} \propto \sum_{i} \left( \begin{matrix}(S_i^x)^2 - (S_i^y)^2\\2S_i^xS_i^y \end{matrix}\right)
\label{eq:SI_nematic}
\end{equation}
as described in Ref.\ \cite{donoway_multimodal_2024} and as depicted in Fig.\ \ref{fig:setup_corot_timedep}b. From Eq.~\eqref{eq:SI_nematic} it is also simple to understand that overall rotations of the spin structure correspond to changes in $\theta$, whereas relative motion between spins will in general change the $\eta$.

\begin{figure*}[t!]
    \centering
    \includegraphics[scale=0.9]{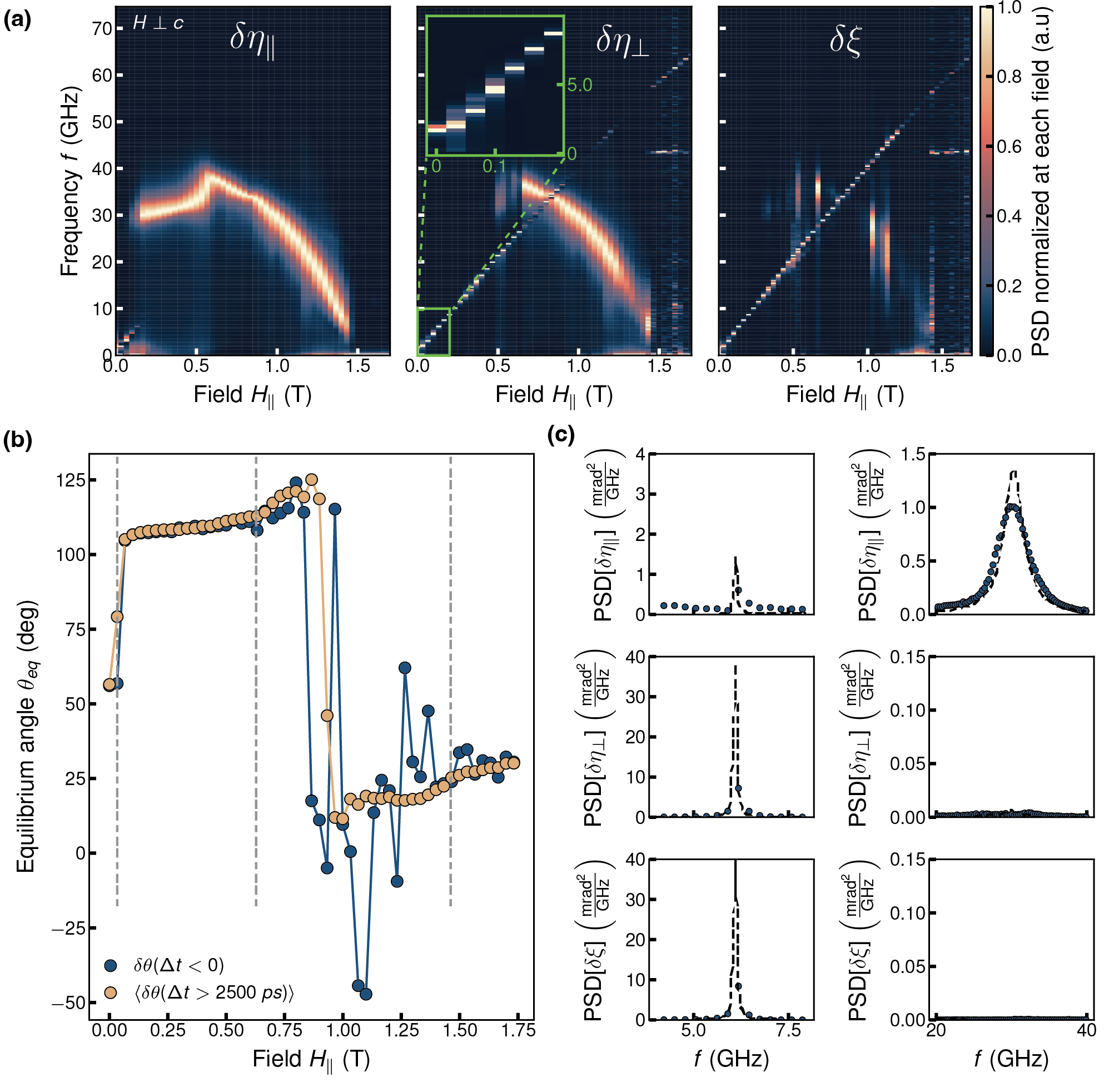}
    \caption{\textbf{(a)} PSD of pump-induced spin-waves in $\delta\el$, $\delta\et$, and $\delta\xi$ channels as a function of in-plane magnetic field ($H\perp\bm{c}$) and frequency, normalized at each field for ease of viewing. Inset: several features at the Goldstone mode frequency track with each other as a function of field. \textbf{(b)} Equilibrium nematic angle $\theta_\text{eq}$ as a function of field as computed by both the $\Delta t < 0$ signal and as the average angle at $\Delta t > \unit[2500]{ps}$. Dashed lines indicate $H_\text{f}$, $H_\text{hf}$, and $H_\text{sat}$. \textbf{(c)} Example power spectra and corresponding fits at $H \approx \unit[0.13]{T}$. Rows from top to bottom correspond to $\delta\el$, $\delta\et$, and $\delta\xi$ channels respectively, the first column shows the Goldstone mode, and the second column shows the much broader optical mode.}
    \label{fig:SI_in_plane} 
\end{figure*}

For an initial linear polarization angle $\phi$, the final polarization state of the probe is given by

\begin{equation}
\left(\begin{matrix}E_H\\E_V\end{matrix}\right) = J_\text{HWP}(\phi/2 + 22.5^\circ) \cdot J_\text{samp} \cdot \left(\begin{matrix}\sin\phi\\\cos\phi\end{matrix}\right).
    \label{eq:SI_polarization_final}
\end{equation}
The resulting signal $S$ can be calculated using Eq.~\eqref{eq:SI_sig}. Physically, $S$ is directly proportional to $\delta\phi$, the angle by which the polarization vector of incident light rotates upon reflection. To see this, one has to compare $S$ to $\delta\phi$ as extracted from the polarization state immediately after reflection
\begin{equation}
A\left(\begin{matrix}\sin(\phi + \delta\phi) \\ \cos(\phi + \delta\phi) \end{matrix}\right) = J_\text{samp}\cdot \left(\begin{matrix}\sin\phi\\\cos\phi\end{matrix}\right),
\end{equation}
where $A$ is the final amplitude of the light. Solving for $\delta\phi$ and comparing to $S$, at first order, yields that $\delta\phi \approx S/2r^2$.

While $S$ describes the instantaneous signal that reaches the lock-in amplifier in a time-dependent experiment, the modulated signal $\tilde{S}$ only picks out components which oscillate at the reference frequency. At a time delay $\Delta t$ after the pump, $r(\Delta t) = r_0 + \delta r(\Delta t)$, $\bm{\eta}(\Delta t) = \bm{\eta}_0 + \delta\bm{\eta}(\Delta t)$, and $\xi(\Delta t) = \xi_0 + \delta\xi(\Delta t)$. To obtain the modulated signal we consider only the time-dependent terms in $S$, yielding the lock-in signal,

\begin{equation}
    \tilde{S}(\phi, \Delta t) = \tilde{A}\sin(2(\phi - \tilde{\Theta})) + \tilde{B}\sin(4(\phi - \tilde{\Phi})) + \tilde{C},
    \label{eq:SI_photovoltage_mod}
\end{equation}

\noindent
where,

\begin{widetext}
\begin{align}
    \tilde{A}(\Delta t) &= 2\sqrt{(r_0\delta\eta_x + \eta_{0,x}\delta r + \delta r \delta\eta_x)^2 + (r_0\delta\eta_y + \eta_{0,y}\delta r + \delta r \delta\eta_y)^2}\label{eq:SI_corot_mod_A},\\
    \tilde{\Theta}(\Delta t) &= \frac{1}{2}\tan^{-1}\left(\frac{r_0\delta\eta_y + \delta r\eta_{0,y} + \delta r\delta\eta_y}{r_0\delta\eta_x + \delta r\eta_{0,x} + \delta r\delta\eta_x}\right),\label{eq:SI_corot_mod_0}\\
    \tilde{B}(\Delta t) &= \sqrt{(\delta\eta_x\eta_y + \delta\eta_y\eta_x + \delta\eta_x\delta\eta_y)^2 + (\delta\eta_y\eta_y - \delta\eta_x\eta_x + \delta\eta_y^2 - \delta\eta_x^2)^2},\label{eq:SI_corot_mod_B}\\
    \tilde{\Phi}(\Delta t) &= \frac{1}{4}\tan^{-1}\left(\frac{\delta\eta_x\eta_y + \delta\eta_y\eta_x + \delta\eta_x\delta\eta_y}{\delta\eta_y\eta_y - \delta\eta_x\eta_x + \delta\eta_y^2 - \delta\eta_x^2}\right)\label{eq:SI_corot_mod_1}\\
    \tilde{C}(\Delta t) &= 2(r_0\delta\xi + \xi_0\delta r + \delta r \delta\xi),\label{eq:SI_corot_mod_C}
\end{align}
\end{widetext}

\noindent
We expect $r_0>>\eta_{0,x}, \eta_{0,y}, \xi, \delta r, \delta\eta_x, \delta\eta_y, \delta\xi$, and so $\tilde{A}, \tilde{C} >> \tilde{B}$. We can approximate the coefficients in the first and last term of Eq.~\eqref{eq:SI_photovoltage_mod} to first order as

\begin{align}
    \tilde{A} &\approx 2r_0\sqrt{\delta\eta_x^2 + \delta\eta_y^2} = 2r_0\delta\eta\\
    \tilde{\Theta} &\approx \frac{1}{2}\tan^{-1}\left(\frac{\delta\eta_y}{\delta\eta_x}\right) = \delta\theta \\
    \tilde{C} &\approx 2r_0\delta\xi.
\end{align}

\noindent
A measurement of the balanced signal as a function of probe polarization $\phi$ is thus a measurement on $\delta\eta$ and $\delta\theta$. Noting that $r\approx r_0$, $\delta\phi \approx \tilde{S}/2|r_0|^2$, and we arrive at Eq.\ \ref{eq:corotations}. In so doing, we redefine $\delta\eta/r_0 \rightarrow \delta\eta$ and $\delta\xi/r_0 \rightarrow \delta\xi$ such that $\delta\eta$ and $\delta\xi$ are in dimensions of radians. Note that since $\delta\phi$ is proportional to the ratio of two measured intensities ($\tilde{S}$ and $|r_0|^2 \approx |E_V|^2 + |E_H|^2)$), we need not measure intensities in absolute units. We also note that the data collected for $\delta\phi(\phi)$ is fit to Eq.~\eqref{eq:corotations} plus a term like $\tilde{B}\sin(4(\phi + \tilde{\Phi}))$, as in Eq.~\eqref{eq:SI_photovoltage_mod}, however only the $\delta\eta\sin(2(\phi - \delta\theta))$ term is used in the subsequent analysis. As Fig.~\ref{fig:setup_corot_timedep}e shows, the $4\phi$  term is insignificant at lower fields, however it becomes more prominent at the higher fields.

\begin{figure}[t!]
    \centering
    \includegraphics[scale=0.9]{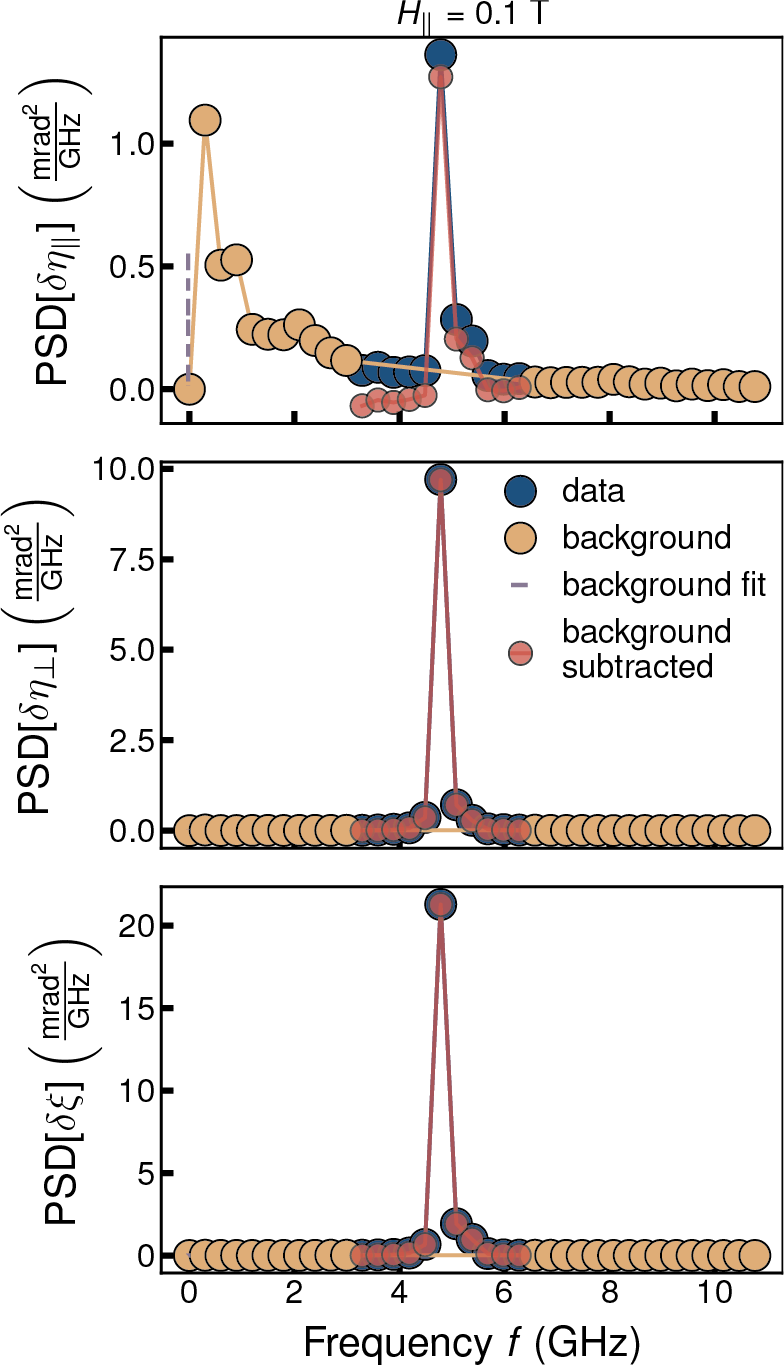}
    \caption{Background subtraction of power spectra from position A for isolation of Goldstone mode spectral weight in $\delta\el$, $\delta\et$, and $\delta\xi$ channels at $H_\parallel=\unit[0.1]{T}$.}
    \label{fig:SI_spectral_weight} 
\end{figure}

The analysis presented above is for an ideal setup with no birefringence of its own. For our purposes, the most important artifact introduced by the optical system comes from cross terms in Eq.~\eqref{eq:SI_corot_mod_A}-\ref{eq:SI_corot_mod_C} which couple the stead-state birefringence terms $\eta_{0,x}$ and $\eta_{0,y}$ to $\delta r$. The steady-state birefringence is dominated by the setup, and we mitigate its effect by ensuring that at each value of $\phi$ and field $H_\parallel$ the photodiodes are \textit{balanced} at DC. To achieve this, we tune $\varphi$ (instead of setting it to $\varphi = \phi/2 + 22.5^\circ$ as in Eq.~\eqref{eq:SI_polarization_final}) at each measurement setting such that $S=0$ when $f_\text{ref}=0$. At high magnetic fields, the cross term $\xi_0\delta r$ in $\tilde{C}$ can also be a problem because of the Faraday rotation in the windows of the cryostat; balancing helps mitigate the impact of such terms as well. For a further analysis of the kinds of artifacts that can occur in polarimetry experiments, see Ref.~\cite{sunko_spin-carrier_2023}.

\section{Analysis of time-resolved data and extraction of mode frequencies}\label{sec:SI_in_plane}

The raw data for $\delta\phi(\phi, \Delta t)$ collected via the lock-in technique described above passes through several layers of analysis; the the time dependence is decomposed into $\delta\el$, $\delta\et$ and $\delta\xi$ channels, spin-wave frequencies as a function of magnetic field are extracted, and the amplitude of oscillations at each frequency in each channel is determined. Here, we explain these analyses in detail and present additional visualizations of the data.

As explained in the main text, Eq.~\eqref{eq:corotations} is used to obtain $\delta\bm{\eta}(\Delta t)$ and $\delta\xi(\Delta t)$ from $\delta\phi(\phi, \Delta t)$. The next step is to decompose $\delta\bm{\eta}$ into $\delta\el$ and $\delta\et$, the components which are longitudinal and transverse to the equilibrium nematic director.  To accomplish this, we must evaluate $\theta_\text{eq}$, the equilibrium nematic angle in the lab frame, from the signal at $\Delta t <0$. The nonzero signal at $\Delta t < 0$ indicates that the system does not relax to thermal equilibrium in between pump pulses, which are spaced about $\unit[3.3]{\mu s}$ apart. In this ``hot" regime, where coherent dynamics are damped out but the system is still cooling, the order parameter is simply reduced in amplitude ($\delta\bm{\eta} \parallel \bm{\eta}$), an assertion that is supported by the temperature independence of $\delta\theta$ \cite{donoway_multimodal_2024}. Consequently, $\delta\theta$ at negative times is equal to $\theta_\text{eq}$. Fig.~\ref{fig:SI_in_plane}b shows $\theta_\text{eq}$ as a function of magnetic field. We also show $\theta_\text{eq}$ as evaluated by taking the average of $\delta\theta$ for $\Delta t > \unit[2500]{ps}$. The two methods agree well, which suggests that the non-oscillatory contribution to $\delta\bm{\eta}(t)$ corresponds to a reduction of the nematic amplitude in both time regimes.

Both methods clearly show the spin-flop transition around $H_\text{f} = \unit[0.035]{T}$, followed by a kink at $H_\text{hf} = \unit[0.63]{T}$ that indicates the helix-fan transition. The latter method is more robust beyond $H_\text{hf}$, where the signal at $\Delta t < 0$ is too weak to get a reliable measurement of $\delta\theta$. In this regime, saturation of the magnetization is observed at $H_\text{sat} = \unit[1.46]{T}$, where the optical mode reaches a frequency minimum (see Fig.~\ref{fig:field_dependence}), which is a generic feature of magnetic systems at saturation, and after which $\theta_\text{eq}$ appears to reach a plateau.

For consistency, we use the former method to extract $\theta_\text{eq}$ (found to yield clearer results at low fields where the negative time signal is strong) and use it to decompose $\delta\bm{\eta}$ into $\delta\el$ and $\delta\et$. The exception is for the maps shown in Fig.\ \ref{fig:SI_in_plane}a, for which the latter method is used to more accurately capture the behavior past the helix-fan transition.

The resulting time traces for $\delta\el(\Delta t)$, $\delta\et(\Delta t)$, and $\delta\xi(\Delta t)$ are collected as a function of magnetic field. For each field, they are subtracted from a constant offset and transformed into the frequency domain via a FFT for $\Delta t > 0$. This produces power spectral density (PSD) maps over frequency and field such as in Fig.~\ref{fig:SI_in_plane}a, which shows the PSD in the three channels for fields applied in the easy-plane at position A. The field evolution of the two modes is evident from these maps without further analysis.

\begin{figure*}[t!]
    \centering
    \includegraphics[scale=0.9]{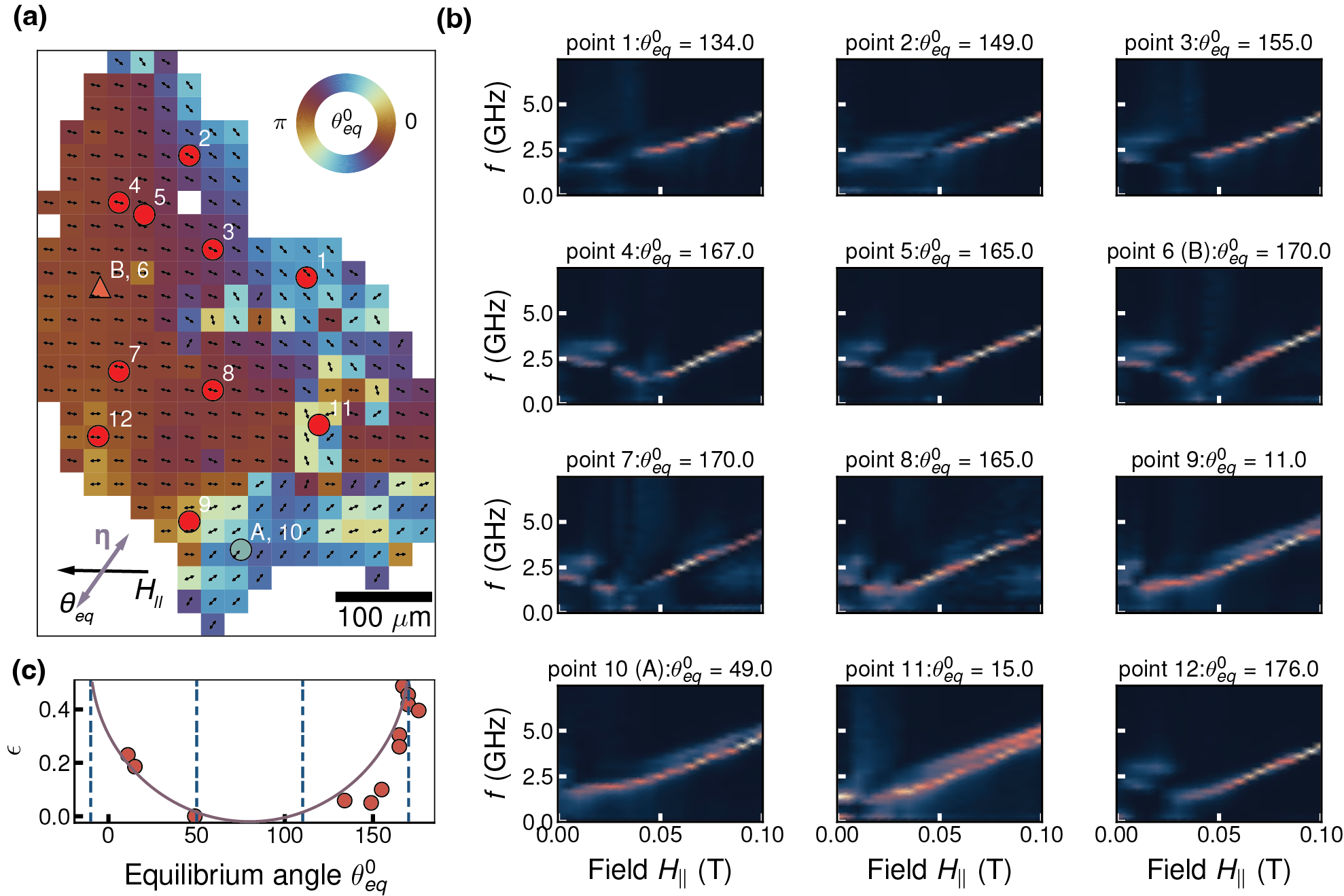}
    \caption{\textbf{(a)} Map of equilibrium nematic angle $\theta_\text{eq}^0$ across the sample at zero applied field, with various positions labeled; positions 10 and 6 are near A and B from the main text. \textbf{(b)} PSD as a function of in-plane field and frequency at each position, normalized at each field, where the underlying time traces are taken at a single probe polarization angle $\phi = 90^\circ$. \textbf{(c)} Relative difference between zero field frequency $f_0$ and minimum frequency $f_\text{min}$, $\epsilon = (f_0 - f_\text{min})/f_0$, a metric for quantifying the dip in frequency near the spin-flop transition, as a function $\theta_\text{eq}$ at zero field. The distribution is peaked near $\theta_\text{eq} = 0^\circ$ and $180^\circ$, suggesting that the anisotropy is 2-fold symmetric. Dotted lines show where peaks in the distribution are expected for 6-fold symmetry, and the solid line is a guide to the eye.}
    \label{fig:SI_lowfield} 
\end{figure*}

Note that the feature around $\unit[40]{GHz}$ that appears at high field and is field-independent is an artifact of the measurement (due to some steady source of laboratory noise) which is present at all field. This was verified by the fact that it persists above $T_{N1}$ and the observed frequency scales with the integration time at each time-step. It is only prominent in the normalized maps at these fields because all other signals are extremely small.

Several approaches are possible to get mode frequencies from FFTs; (1) using a peak-finding algorithm such as the Scipy function \verb|scipy.signal.find_peaks|, (2) fitting the FFTs to a model. In practice, we use a combination of both, whereby method (1) is used to approximately find peaks, and then method (2) is used for refinements. For the model, we take each mode to be a damped, driven harmonic oscillator with the following functional form:

\begin{align}
    \label{eq:SI_peakfitting_time}
    A(t) &= \Theta(t)A_0e^{-\Gamma t}\cos({2\pi f_0 t + \varphi}) \\
    \label{eq:SI_peakfitting_freq}
    \tilde{A}(f) &= \text{FFT}[A(t)] = \frac{1}{2}\frac{A_0 e^{i\varphi}}{(\Gamma + 2\pi i(f-f_0))}
\end{align}

\noindent
where $\Theta(t)$ is the Heaviside step function, $f_0$ is natural resonance frequency of the mode, $A_0$ is the amplitude, $\Gamma$ is the damping, and $\varphi$ is the phase. 

Furthermore, since the same set of modes give rise to oscillations in the three measurement channels, we require that the mode frequency and damping be the same across all three channels for a given mode. The mode frequencies are thus extracted by simultaneously fitting the FFTs at each field for two modes, where only the amplitude and phase are allowed to vary across channels. Fig.~\ref{fig:SI_in_plane}c shows an example of the fits obtained at $H\approx\unit[0.13]{T}$, where the first column is a close up view of the Goldstone mode, whereas the second column focuses on the optical mode. We emphasize that the purpose of using a fit model is to obtain a controlled estimate of the mode frequency, rather than as an attempt to fully capture the data.

Lastly, getting an estimate of the amplitude of oscillation for a given mode in each channel requires some care, especially for the Goldstone mode. As can be seen in Fig.~\ref{fig:SI_in_plane}c, the density of frequency data points is sparse compared to the sharp linewidth of the mode, which allows the fits to vary widely in amplitude. In addition, the fit amplitude is often influenced by the background, which is hard to incorporate into the model. Instead, we opt for a model independent approach, demonstrated in Fig.~\ref{fig:SI_spectral_weight}. In this approach, a width about the center frequency is defined as ``peak data" and points outside of this width are considered ``background data." The background data is fit to a third-order polynomial, interpolated over the data, and then subtracted to yield a background subtracted PSD, which is then integrated to obtain the spectral weight of the peak shown in Fig.~\ref{fig:strain}e.

\begin{figure*}[t!]
    \centering
    \includegraphics[scale=0.9]{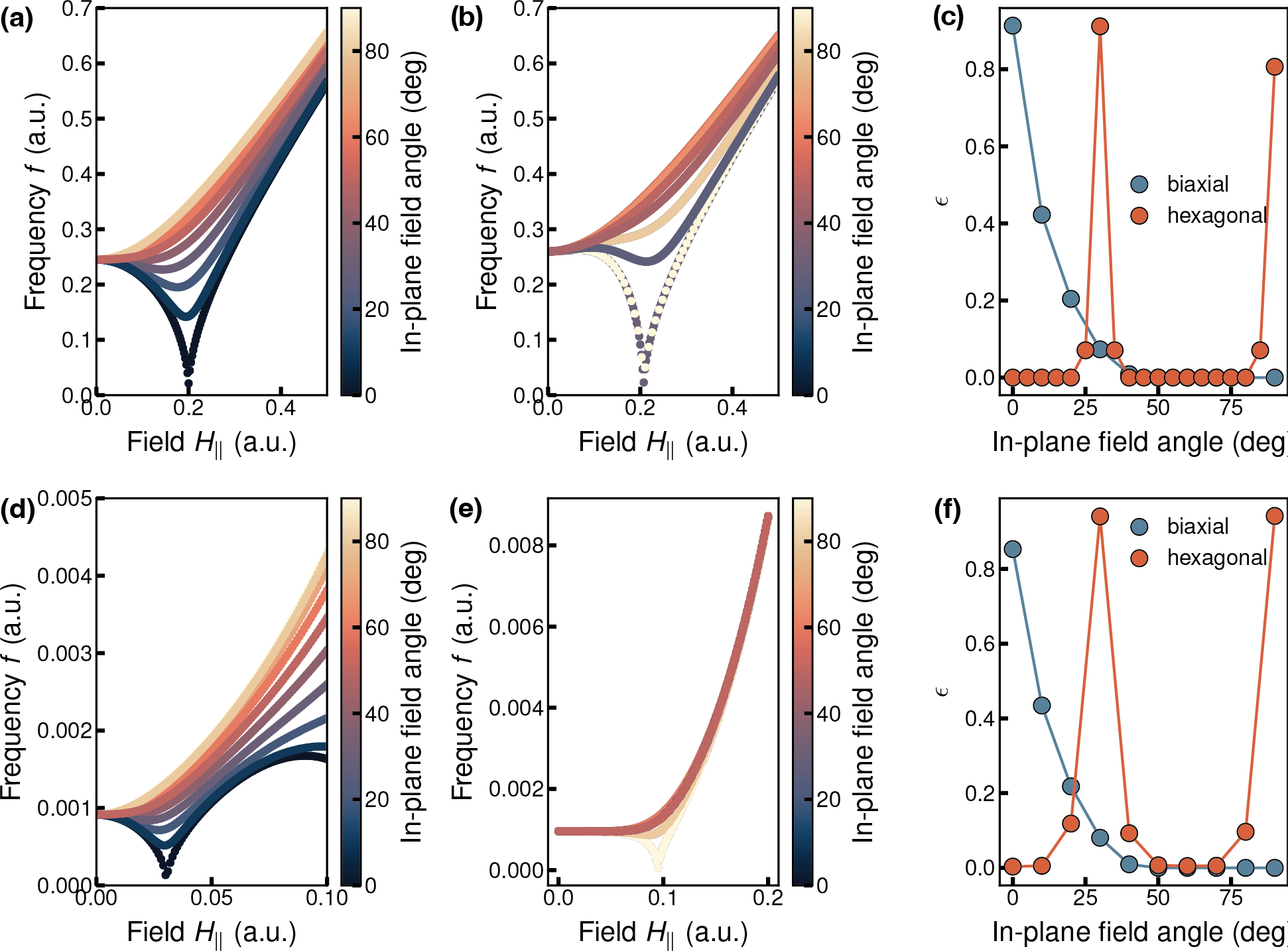}
    \caption{Linear spin-wave theory calculations of the field dependence of the Goldstone mode frequency for different spin models in the presence of weak in-plane anisotropy. Here we show results at low fields relative to saturation for various in-plane field orientations for \textbf{(a)} easy-plane  biaxial antiferromagnet ($J=-1$, $K_z=-1$, $K_x = 0.005$ in Eq.\ \ref{eq:SI_afm}), \textbf{(b)} easy-plane antiferromagnet with hexagonal anisotropy ($J=-1$, $K_z = -1$, $K_\text{hex} = 3.125\times 10^{-4}$ in Eq.\ \ref{eq:SI_afm}), \textbf{(d)} easy-plane biaxial j1-j2 helix ($J_1=1$, $K_z=-1$,  $K_x = 0.005$ in Eq.\ \ref{eq:SI_j1j2}), and \textbf{(e)} easy-plane j1-j2 helix with hexagonal anisotropy ($J_1= 1$, $K_z=-1$, $K_\text{hex} = 1.02\times 10^{-8}$ in Eq.\ \ref{eq:SI_j1j2}). \textbf{(c)} and \textbf{(f)} show $\epsilon$ vs the applied field angle for the two types of anisotropy in the antiferromagnet and helix respectively. For all calculations, $0^\circ$ corresponds to alignment of the field along one of the easy-axes.}
    \label{fig:SI_spin_models} 
\end{figure*}

\subsection{Beat frequencies}

The slight beating in the envelopes of $\delta\et$ and $\delta\xi$ noted in the main text can be seen as a satellite peak of the Goldstone mode (inset of Fig.~\ref{fig:SI_in_plane}a, middle panel). Several possibilities exist to explain this feature. (1) As the broken helix, with six spins in a magnetic unit cell, is expected to have six spin-wave modes, one possibility is that there are simply two closely spaced modes. However, that the two frequencies track each other with the applied field makes this unlikely. (2) Our measurement integrates over excitations with momentum up to $k_\text{max} \approx \unit[0.2]{\mu m^{-1}}$, as determined by the $\approx\unit[5]{\mu m}$ probe spot size. This additional frequency could therefore arise from a peak in the spin-wave density of states at small, nonzero $k < k_\text{max}$ induced by the magnetostatic interaction, as observed in CrSBr \cite{sun_dipolar_2024}. (3) Another possibility is that the beating is a result of spin-wave propagation out of the probe volume, also induced by the magnetostatic interaction. Magnetostatic interactions can lead to a linear dispersion near $\bm{k}=0$ even in a gapped system and hence a nonzero group velocity.

\section{Additional spatially resolved spin-wave frequencies at low field}\label{sec:SI_lowfield}

To demonstrate that the spin-wave frequency at zero field for the broken helix is determined by strain within the easy-plane, we measured the pump-induced polarization rotation $\delta\phi(\Delta t)$ at a variety of positions across the sample at a probe polarization of $\phi=90^\circ$ as a function of field (Fig.~\ref{fig:SI_lowfield}a shows a map of the equilibrium angle across the sample at zero field, and marks the positions at which we measured field dependent frequencies). Although probing at a single polarization angle does not enable a decomposition into $\delta\el, \delta\et$, and $\delta\xi$, we are still able to visualize the evolution of the Goldstone mode frequency with field by plotting the power spectra (Fig.~\ref{fig:SI_lowfield}b). Evidently the trend described in Fig.~\ref{fig:strain}c is systematic, where a strain orientation-dependent dip in frequency at the spin-flop transition is observed.

We can quantify this trend by defining a metric for the prominence of the dip
\begin{equation}
    \epsilon = \frac{f_0 - f_\text{min}}{f_0},
\end{equation}
where $f_0$ is the frequency at zero field and $f_\text{min}$ is the minimum frequency in a field. Fig.~\ref{fig:SI_lowfield}c shows $\epsilon$ plotted against $\theta_\text{eq}$ at each position. The solid line is a guide to the eye. This metric appears to respect two-fold symmetry as expected for a strain-induced easy axis. On the other hand, hexagonal anisotropy would have six-fold symmetry, which is indicated on the plot by the vertical lines. These features are reproduced in simple spin-wave calculations of antiferromagnets and helices below. 

\section{Linear spin-wave theory and symmetry analysis}\label{sec:SI_sw_theory}

Here we discuss the dynamical modes and their relation to the energy costs of static fluctuations. Given a ground state spin configuration is $\bm{S}^0_j$ we can express small fluctuations of the spins as
\begin{align}
    \delta \bm{S}_j = \sum_{a=1,2} \Big(  m^a_j \bm{e}^a_j - \frac{1}{2}[m^a_j]^2 \bm{S}^0_j\Big) + O(m^3),
\end{align}
where for each site $j$, $\bm{e}^1_j$, $\bm{e}^2_j$ and $\bm{S}^0_j$ are a right-handed set of orthonormal vectors $\bm{e}^1_j \times \bm{e}^2_j = \bm{S}^0_j$. The scalars $m^a_j$ are the amplitudes of the fluctuations along $\bm{e}^a_j$.

Up to quadratic order in these fluctuations we can write the energy of the system as
\begin{align}
    E = E_0 + \frac{1}{2}\sum_{jk \atop ab} \rho^{ab}_{jk} m^a_j m^b_k + O(m^3).
    \label{eq:SI_quadexp}
\end{align} 
where $\rho^{ab}_{jk}$ are the elements of the $2N \times 2N$ Hessian matrix (where we interpret e.g. $a$ and $j$ as a row index). The Hessian is real, symmetry, and positive semidefinite. 

The semiclassical commutation relation $[m^1_j, m^2_j]=i$ implies the linearized equations of motion $(d/dt)m^a_j = \epsilon^{ab} \partial E/\partial m^b_j$, where $\epsilon^{01}=-\epsilon^{10}=0$ and $\epsilon^{00}=\epsilon^{11}=0$. In terms of the Hessian we then have
\begin{align}
    \frac{d}{dt}m^a_j = \sum_{bk} \epsilon^{ab} \rho^{ab}_{jk} m^b_j.
\end{align}
The mode matrix is given by $D^{ab}_{jk} = -(i/2\pi) \epsilon^{ab}\rho^{ab}_{jk}$. 

The mode matrix is purely imaginary, and its eigenvalues are the mode frequencies. If $m^a_j$ are the components of an eigenvector with eigenvalue $f$, then by definition $\sum_{bk} D^{ab}_{jk} m^b_k = f m^a_j$. Taking the complex conjugate of this expression we have $-\sum_{bk} D^{ab}_{jk} [m^b_k]^* = -f^* [m^a_j]^*$. Therefore, if $f$ is an eigenvalue, then so is $-f^*$. For a stable ground state configuration all frequencies are real, and they therefore appear in oppositely signed pairs $\pm f$.

The symmetry constraint on the characteristic polynomial $p(f,\bm{H})$, introduced in the main text, is
\begin{equation}
p(f, \bm{H}) = p(f, R\bm{H}),
\end{equation}
for all $R$ in the point group $G$ of the magnetic ground state.\cat

\begin{figure*}[t!]
    \centering
    \includegraphics[scale=0.9]{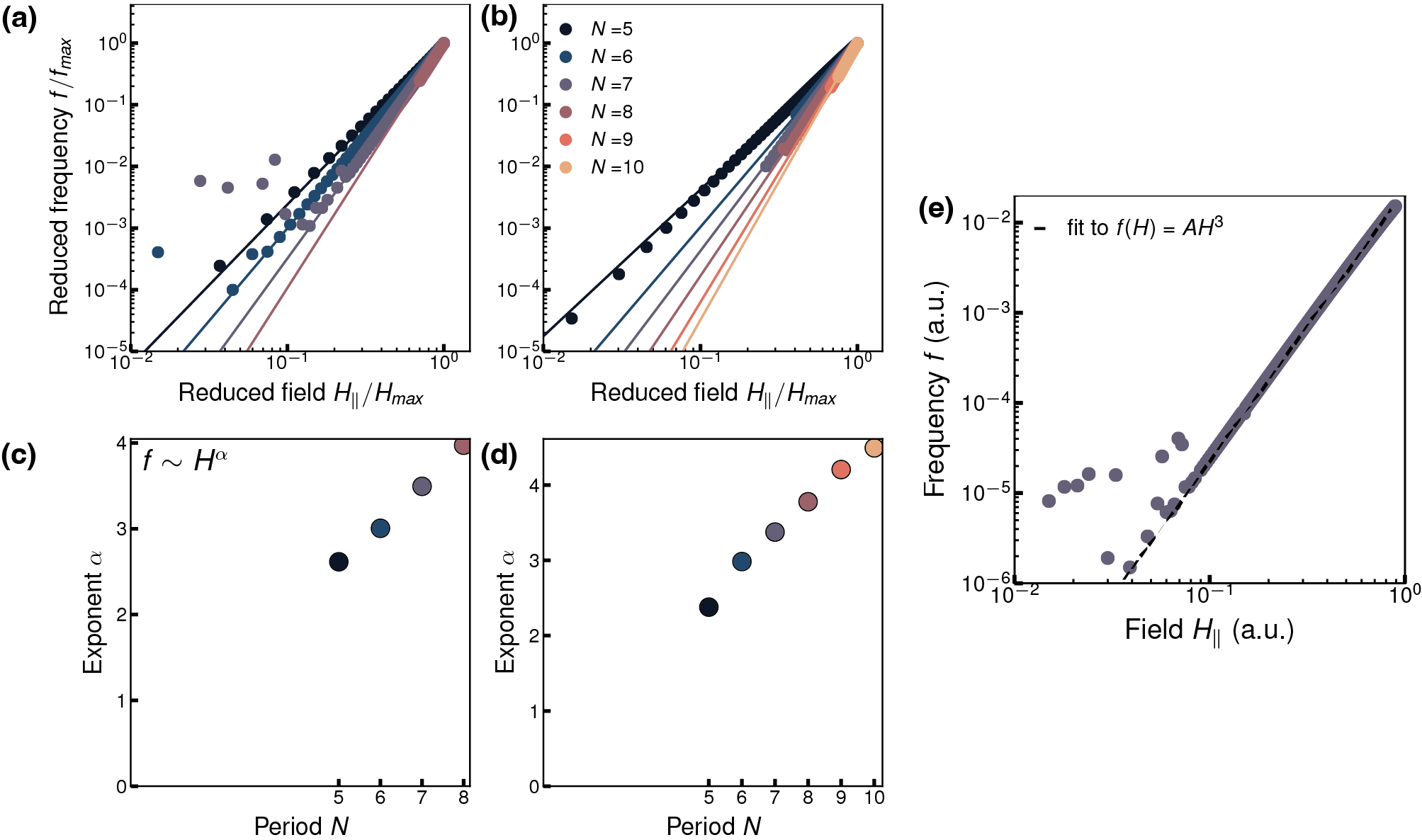}
    \caption{Logarithmic plots of frequency versus in-plane field and power law fits to $f(H) = AH^\alpha$ for different $N$-helices for \textbf{(a)} j1-j2 models with $J_1=1, J_2=-|J_1|/(4\cos(2\pi/N)), K_z=-1$, and \textbf{(b)} RKKY-like models with $C_1=C_2=0$. Since $A$ strongly depends on $N$, we plot in reduced variables for comparison. Scatter at low fields is caused by numerical error in relaxing the ground state configurations at each field. The extracted exponents $\alpha$ versus $N$ are shown in \textbf{(c)} for the j1-j2 models, and \textbf{(d)} for the RKKY-like models, demonstrating that $\alpha(N) = N/2$. \textbf{(e)} Logarithmic plot of frequency versus field of a 6-fold RKKY helix with biquadratic exchange terms $C_1=C_2=-0.5$, showing a clear $H^3$ field dependence.}
    \label{fig:SI_helix_powerlaws} 
\end{figure*}

\section{Lowest order contributions to Goldstone mode frequency}\label{sec:SI_theory}

In this section we discuss the conditions under which the Goldstone mode frequency can scale as $\omega \sim H^{\alpha}$ with $\alpha > 1$, focusing on easy-plane magnets with $\text{U}(1)$ symmetry. In this setting we can understand the Goldstone mode as arising from two conjugate coordinates: smooth rotations of the spins within the plane, and fluctuations of the magnetization out of the plane. While the energy cost of a smooth rotation vanishes at long wavelengths, the energy cost of an out-of-plane magnetization fluctuation does not. As a consequence, in a symmetry-breaking external field, the scaling of the Goldstone mode gap with field strength is controlled by the `stiffness' associated with the coordinate which, in the zero field limit, becomes a global rotation. We can therefore focus on calculating this stiffness, which corresponds to the lowest eigenvalue of the Hessian characterizing the energy costs of spin fluctuations. 

In an easy-plane magnet the $2N \times 2N$ Hessian matrix, having elements $\rho^{ab}_{jk}$, consists of two $N \times N$ blocks. One of these blocks, which we denote $\hat \rho$ with elements $\hat \rho_{jk}$, describes the energy costs of in-plane fluctuations, and in zero field one of the eigenvalues of $\hat \rho$ is zero. If we adopt the convention that $\bm{e}^1_j$ is in the easy plane, then $\hat \rho_{jk}=\rho^{11}_{jk}$. The other block, with elements $\rho^{22}_{jk}$ describes the energy costs of out-of-plane fluctuations; all eigenvalues of that block are finite even in zero field. Since our focus will be on $\hat \rho$, when describing in-plane fluctuations we will omit superscripts, i.e. $m_j = m^1_j$ in this section. 

For a classical spin system with easy-plane anisotropy we write $\bm{S}_j = (\cos \theta_j,\sin\theta_j,0)$ for spin $j$. With rotational symmetry and Heisenberg interactions, the energy takes the general form
\begin{align}
	E = -\sum_{j<k} J_{jk} \cos[\theta_j - \theta_k] - H \sum_j \cos[\theta_j],
\end{align}
where we have included an in-plane magnetic field $B$. In a system with $N$ spins per unit cell we can understand zero momentum fluctuations by considering just $N$ spins with periodic boundary conditions. Making this restriction, the index $j=0,\ldots,(N-1)$, while $E$ corresponds to the energy per unit cell. The ground state satisfies $\partial E/\partial \theta_j=0$ for all $j$, i.e. 
\begin{align}
	\sum_k J_{jk} \sin[\theta_j - \theta_k]  + H \sin \theta_j = 0.
\end{align}
For small $H$ the solution to this equation can be expressed as $\theta_j = \theta_j(H) = \sum_{k=0} H^k \theta^{(k)}_j$; we denote by $\delta \theta_j = \sum_{k=1} H^k \theta^{(k)}_j$ the terms which vanish as $H \to 0$. Our focus will be on systems having zero magnetization for $H=0$, i.e. $\sum_j \cos [\theta^{(0)}_j] = 0$. To determine $\theta^{(1)}_j$, let us write the ground state energy as $E = \sum_k H^k E^{(k)}$ with $E^{(1)}=-H \sum_j \cos[\theta^{(0)}_j]$ and
\begin{align}
	E^{(2)} =& \frac{1}{2} \sum_{j<k} J_{jk} \cos[\theta^{(0)}_j - \theta_k^{(0)}](\theta_j^{(1)} - \theta_k^{(1)})^2\notag \\
    &+ \sum_j \sin[\theta^{(0)}_j]\theta^{(1)}_j.
\end{align}
The first line can be expressed as $\frac{1}{2}\sum_{jk} \hat \rho_{jk} \theta^{(1)}_j \theta^{(1)}_k$ where $\hat \rho_{jk}$ is the Hessian matrix for the zero-field system. This matrix is positive semidefinite, real and symmetric. In the presence of a field we then find the first-order change of the ground state configuration $\theta^{(1)}_j = -\sum_k\hat\rho^{-1}_{jk} \sin[\theta_k^{(0)}]$. In the case where we have a Goldstone fluctuation at $H=0$ the matrix $\hat\rho_{jk}$ has a zero eigenvalue; the condition that $\theta^{(1)}_j$ is finite then forces us to select zero-field ground states such that the vector with components $\sin[\theta_k^{(0)}]$ has no overlap with the zero eigenvector of $\hat\rho_{jk}$.

Our aim is to identify when there exist fluctuations around the ground state configuration having energy costs parametrically smaller than $\sim H^2$, which is the lowest order allowed by symmetry in a system with zero net magnetic moment. Quite generally, in-plane fluctuations $\theta_j \to \theta_j + m_j$ around the ground state $\theta_j = \theta_j(H)$ are described at second order by
\begin{align}
	\delta E &= \frac{1}{2} \sum_{j<k} J_{jk} \cos[\theta_j-\theta_k](m_j-m_k)^2 \label{eq:deltaE} \\&+ \frac{1}{2}H \sum_j \cos[\theta_j] m_j^2, \notag
\end{align}
note that linear terms cancel since we are considering an energy minimum. Expanding the configuration $\theta_j$ in powers of $H$, the above expression defines the Hessian which characterizes fluctuations around the energy minimum at $H \neq 0$. The basic question is whether we can find a useful upper bound on the lowest eigenvalue $\lambda_0(H)$ of the $H \neq 0$ Hessian. 

We can find a variational upper bound on the lowest Hessian eigenvalue via $\lambda_0(H) \leq \delta E / \bm{m}^T\bm{m}$, where $\delta E$ is given by Eq.~\eqref{eq:deltaE} and $\bm{m}$ in a $N$-component vector with elements $m_j$. For $H=0$, there is a zero energy fluctuation $m_j = m$, i.e. a global rotation. For $H \neq 0$ we use the ansatz $m_j = m + H\chi_j + O(H^3)$ with $\sum_j \chi_j = 0$. Inserting this ansatz into $\delta E$,  and working to second order in $H$, we are left with
\begin{align}
	&\delta E = \frac{1}{2} H^2 \sum_{j<k} J_{jk} \cos[\theta^{(0)}_j - \theta^{(0)}_k] (\chi_j - \chi_k)^2 \\&+ H^2 m \sum_j \cos[\theta^{(0)}_j] \chi_j - \frac{1}{2}H^2 m^2 \sum_j \sin[\theta^{(0)}_j] \theta^{(1)}_j + O(H^3). \notag
\end{align}
This expression depends only on the first order correction to the ground state configuration $\theta^{(1)}_j = -\sum_k\hat\rho^{-1}_{jk} \sin[\theta_k^{(0)}]$. We have also used $\sum_j \cos[\theta^{(0)}_j]=0$. We then find
\begin{align}
	\delta E &= \frac{1}{2}H^2 \bm{\chi}^T \hat\rho \bm{\chi} + H^2 m \bm{c}^T \bm{\chi}\\ &+ \frac{1}{2} H^2 m^2 \bm{s}^T \hat \rho^{-1}\bm{s} + O(H^3).\notag\end{align}
Here we have defined the $N$-component vectors $\bm{s}$ and $\bm{c}$, whose respective entries are $\sin [\theta^{(0)}_j]$ and $\cos [\theta^{(0)}_j]$, as well as the $N$-component vector $\bm{\chi}$ with entries $\chi_j$. 

Next we minimize $\delta E$ with respect to $\chi_j$ subject to the constraint $\sum_j \chi_j=0$, i.e. we minimize $\delta E-\mu\sum_j \chi_j$ where $\mu$ is a Lagrange multiplier. This gives
\begin{align}
	\rho \bm{\chi} + m \bm{c} - \mu = 0.
\end{align}
Recall that $\sum_j c_j=0$: $\bm{c}$ has zero overlap with the vector describing a uniform rotation. Therefore, $\rho^{-1} \bm{c}$ is finite, and $\sum_j [\hat\rho^{-1}\bm{c}]_j=0$. This means that we can set $\mu=0$, with $\bm{\chi} = -m \rho^{-1} \bm{c}$ satisfying the constraint $\sum_j \chi_j =0$. From this we find
\begin{align}
	\delta E = \frac{1}{2}(H m)^2\big( -\bm{c}^T \hat\rho^{-1} \bm{c} + \bm{s}^T \hat\rho^{-1} \bm{s}\big) + O(H^3).
\end{align}
To turn this into a bound on the lowest eigenvalue $\lambda_0(H)$ of the Hessian of the $H \neq 0$ system, we need to normalize by $\bm{m}^T \bm{m} = N m^2$. The result is 
\begin{align}
	\lambda_0(H) \leq \frac{H^2}{2N}\big(  \bm{s}^T \hat\rho^{-1} \bm{s} - \bm{c}^T \hat\rho^{-1} \bm{c}\big) + O(H^3).
\end{align}
The dependence of the Goldstone mode frequency on field can be determined from $\omega \sim \lambda^{1/2}$, which holds at the lowest nonzero order in $H$. We have therefore arrived at a general condition for the vanishing of $O(H)$ contribution to the Goldstone mode frequency. Note also that, if the $O(H^2)$ term in this expression vanishes, and if the zero-field ground state is symmetric under a $\pi$-rotation within the plane, the lowest order contribution to $\lambda$ is $O(H^4)$, and hence $\omega = O(H^2)$.

Now consider a system where the ground state breaks the discrete symmetry of the Hamiltonian under translations by individual lattice spacings to a discrete symmetry under translations by $N$ lattice spacings. The matrix $\rho$ can then be diagonalized in a plane wave basis. We define the complex vectors $\bm{v}^n$, with $n=0,\ldots,N-1$, via their components $v^n_j =N^{1/2}e^{-i2\pi nj/N}$, such that the zero field Hessian takes the form $\hat\rho = \sum_n \lambda_n(0) v^n_j (v^n_j)^{\dag}$, where $\lambda_n(0)$ are the eigenvalues of the zero field Hessian. Our bound then becomes
\begin{align}
    \lambda_0(H) \leq \frac{H^2}{2N}\sum_{n=1}^{N-1}\lambda_n(0)\Big( |\bm{v}^n_j \cdot \bm{s}|^2 - |\bm{v}^n_j \cdot \bm{c}|^2\Big) + O(H^3). \notag
\end{align}   
A prominent example of a system for which this formula applies is a helix. Setting $\theta_j = \pi(2j+1)/N$ for $j=0,\ldots,N-1$, it can be verified that $|\bm{v}^n_j \cdot \bm{s}|^2=|\bm{v}^n_j \cdot \bm{c}|^2$. The lowest order contributions to $\lambda_0(H)$, and hence to the Goldstone mode frequency, vanish. Another example is the broken helix we study in experiment. In that case $|\bm{v}^n_j \cdot \bm{s}|^2 \neq |\bm{v}^n_j \cdot \bm{c}|^2$ for some values of $n$. The lowest order contributions to the Goldstone mode frequency do not vanish: we find  $\omega \sim H$ as in a conventional antiferromagnet. This is precisely the behavior observed in experiment in the field-dominated (rather than strain-dominated) regime. 

\section{Numerical simulations of helical and antiferromagnetic structures}\label{sec:SI_simulation}

To complement the experimental and theoretical work, we have used linear spin-wave theory calculations to (1) better understand the effect of small easy-axis and hexagonal anisotropies on the Goldstone modes of antiferromagnetic and helical easy-plane systems, and (2) explore the sublinear scaling of the Goldstone mode frequency with applied in-plane field for helices. What follows is a sketch of our computational methods.

The essential task of linear spin-wave theory is to find the eigenvalues of the mode matrix $D$, as outlined in Sec.~ \ref{sec:SI_sw_theory}. Given a particular classical spin Hamiltonian $E$, in order to compute $D$ we need to expand $E$ in terms of $\delta\bm{S}_j$, the fluctuations which are transverse to the ground state spin configuration $\bm{S}^0_j$, as in Eq.\ \ref{eq:SI_quadexp}. Typically, we express $E$ as a function of spins in Cartesian coordinates $(x,y,z)$ for each site $j$, ie.\ $\bm{S}_j = (S_j^x, S_j^y, S_j^z)$. For a general non-collinear magnetic structure, in order to carry out the desired expansion it is instead convenient to express spins in the ``local" frame for each site $j$. For a spin in the ``local" frame, which we denote as $\bm{S}_j^\prime$, $\bm{S}_j^{\prime 0} \parallel z^\prime_i$ and the transverse directions $\bm{e}_j^1$ and $\bm{e}_j^2$ can be chosen to lie along $x^\prime_j$ and $y^\prime_j$ respectively. The relation between spins in the Cartesian and ``local" frames is given by\cite{diep_quantum_2022}

\begin{align}
    \label{eq:SI_LLlocalcoords}
    \bm{S}_j^\prime &= \bm{R}_j\bm{S}_j\\
    \bm{R}_j &= R_z(-\phi_j)R_y(-\theta_j) \\
    \theta_j &=\tan^{-1}\left(\sqrt{(S_j^{x0})^2 + (S_j^{y0})^2}/S_j^{z0}\right)\\
    \phi_j &= \tan^{-1}\left(S_j^{y0}/S_j^{x0}\right),
\end{align}
\noindent
where $R_y$ and $R_z$ are rotation matrices for rotations about the $y$ and $z$ axes, and $\theta_j$ and $\phi_j$ are the polar and azimuthal angles of $\bm{S}_j^0$. 

Given a ground state configuration $\bm{S}_j^0$ which minimizes $E$, and after transforming $E$ to the ``local" frame, it is straightforward to expand $E$ in terms of the transverse fluctuations $\delta\bm{S}_j^\prime$ and form the mode matrix $D$. In practice, we begin with a specific spin Hamiltonian $E$ and use Sympy, a Python based symbolic math package, to generate a matrix function $D$ which takes $\{K_l\}$, the set of couplings in $E$, and $\bm{S}_j^0$ as input. Next, we numerically minimize $E$ for a specific set $\{K_l\}$ to obtain $\bm{S}_j^0$ within a single magnetic unit cell. Lastly, we input both $\{K_l\}$ and $\bm{S}_j^0$ into $D$ to generate a numerical matrix and solve for the normal mode frequencies $f(\{K_l\})$ at momentum $k=0$. Periodic boundary conditions are employed as is standard.

\subsection{Goldstone modes of easy-plane antiferromagnets and helices with small in-plane anisotropy}

We first turn our attention to exploring the effects of small in-plane anisotropies of the field dependence of the Goldstone mode in two simple systems, the collinear antiferromagnet and the j1-j2 helix, a helix stabilized by frustrated nearest-\ and next-nearest-neighbor interactions. Let the various anisotropy energies for a spin $\bm{S}_i$ be denoted by

\begin{equation}
\Delta_i = -\left(K_z\left(S_i^z\right)^2 + K_x\left(S_i^x\right)^2 + K_\text{hex}\Delta_{\text{hex},i}\right).
\end{equation}

\noindent
The first two terms describe uniaxial anisotropy in and out of the spin plane, whereas the last term describes hexagonal in-plane anisotropy, where $\Delta_{\text{hex},i} = \cos(6\phi_i) = \text{Re}[(1/2)(S_i^y - iS_i^x)^6 + \text{c.c})]$. A planar antiferromagnet, with two spins in a unit cell, has a simple Heisenberg-like Hamiltonian

\begin{equation}
    E_\text{afm} = -J\bm{S}_1\cdot\bm{S}_2 - \sum_{i=1,2}\left(\mu_B\bm{H}\cdot\bm{S}_i - \Delta_i\right)
    \label{eq:SI_afm}
\end{equation}

\noindent
$\bm{H}$ is the applied field, and we set $J=-1, K_z=-1, \mu_B = 1$. Fig.~\ref{fig:SI_spin_models}a-b show the field dependence of the Goldstone mode frequency for various in-plane angles of the magnet field spanning $0^\circ$ to $90^\circ$ for $K_x=0.005, K_\text{hex}=0$ and $K_x=0, K_\text{hex}=3.125\times 10^{-4}$ respectively. These parameters were chosen such that the zero field gap is comparable in both cases. Both results are reminiscent of the data in Fig.~\ref{fig:strain}c and in Fig.~\ref{fig:SI_lowfield}b, however for the case of hexagonal anisotropy the six equivalent directions have the same field dependence, and the extreme cases coincide with field angles that are $30^\circ$ apart. This can be seen in Fig.\ \ref{fig:SI_spin_models}c, which show $\epsilon = \frac{f_0 - f_\text{min}}{f_0}$ for both types of anisotropy. In contrast, the uniaxial anisotropy reflects a 2-fold symmetry and the extreme cases are for field angles of $0^\circ$ and $90^\circ$, which matches the experiments.

The simplest model for a helix is the well-studied j1-j2 model, which for our purposes takes the form \cite{fishman_normal_2019}

\begin{widetext}
\begin{equation}
    E_\text{j1j2} = -\sum_{i,j=\pm1,k=\pm2}\left(J_1\bm{S}_i\cdot\bm{S}_{i+j} + J_2\bm{S}_i\cdot\bm{S}_{i+k} \right) - \sum_{i=1,2}\left(\mu_B\bm{H}\cdot\bm{S}_i - \Delta_i\right)
    \label{eq:SI_j1j2}
\end{equation}
\end{widetext}

\noindent
We set $K_z=1$, $J_1=1$, and $J_2 = -|J_1|/2$, which can be shown to stabilize a 6-fold helix (6 spins in a period) for $B=0$. Fig.~\ref{fig:SI_spin_models}d-f show the corresponding field dependence of the Goldstone mode frequency for $K_x=0.005, K_\text{hex}=0$ and $K_x=0, K_\text{hex}=1.02\times 10^{-8}$ respectively.

\subsection{Power law frequency dependence on magnetic field in planar helices}

As discussed in the main text and in Sec.~\ref{sec:SI_theory}, easy-plane spin helices are a class of systems for which the Goldstone mode frequency depends sublinearly on an applied in-plane field. Here, we demonstrate this numerically for the case of the j1-j2 model, a long-range Ruderman–Kittel–Kasuya–Yosida (RKKY) type model, as well as for an RKKY model with biquadratic exchange.

Fig.~\ref{fig:SI_helix_powerlaws}a shows a loglog plot of frequency versus applied field for j1j2 models with $N$-fold helical ground states. In all models $J_1=1, J_2=-|J_1|/(4\cos(2\pi/N)), K_z=-1$ and all other anisotropies are 0. Each plot is fitted to the power law $f(H) = AH^\alpha$ at low fields; since $A$ strongly depends on $N$, we plot in reduced variables $f/f_\text{max}$ and $H/H_\text{max}$ for ease of comparison. The scatter at low fields is caused by numerical error in relaxing the ground state configurations. The case of $N=3$ yields a mode which anomalously stays at $f=0$ at all nonzero fields, and a 4-fold helix is not realizable within the j1j2 model.

Fig.~\ref{fig:SI_helix_powerlaws}c shows the extracted exponent $\alpha$ versus $N$, demonstrating that $\alpha(N) = N/2$. While this has been suggested previously, to our knowledge, this is the first numerical demonstration of this result \cite{zaliznyak_modification_1995}. Thus, it contributes to the literature on the normal modes in helical systems \cite{cooper_spin-wave_1963, fishman_normal_2019}. We also note that this result is consistent with the $N\rightarrow\infty$ limit, which is described by the Sine-Gordon equation and which predicts that $\alpha\rightarrow\infty$, a perfectly soft mode.

We now check that this result is not specific to the j1-j2 model, or even to Heisenberg exchange models. We consider a more realistic model whereby spins interact via an RKKY-like long-range Heisenberg interaction, as well as biquadratic exchange terms which can naturally appear in metallic systems \cite{donoway_multimodal_2024}

\begin{widetext}
\begin{align}
E_\text{RKKY} &= -\sum_{i,j} J(q)_{ij}\bm{S}_i\cdot\bm{S}_j - \sum_i \left(C_1(\bm{S}_i\cdot\bm{S}_{i+1})^2 + C_2(\bm{S}_i\cdot\bm{S}_{i+1})(\bm{S}_{i+1}\cdot\bm{S}_{i+2})  \right) - \sum_i\mu_B\bm{H}\cdot\bm{S}_i\\
J(q)_{ij} &= \frac{J_q\cos(2\pi q\bm{r}_{ij})}{q\bm{r}_{ij}},
\end{align}
\end{widetext}

\noindent
where $\bm{r}_{ij}$ is the distance between sites $i$ and $j$. Because in practice we require a cutoff on the range of the interaction to simulate on a finite system, for an $N$-helix, we cutoff the interaction at $N - 1$ spins. For all simulations, we take $J_q=1$ and set $q$ according to Tab.~\ref{tab:SI_qparams}, chosen such that $E_\text{RKKY}$ is minimized by each $N$-helix. Fig.~\ref{fig:SI_helix_powerlaws}b shows the field dependence of the Goldstone mode frequency for this model with $C_1 = C_2  = 0$, and Fig.~\ref{fig:SI_helix_powerlaws}d shows the associated extracted exponents.

\begin{table}[h]
\centering
\begin{tabular}{c|c}
$N$ & $q$  \\
\hline
5 & 0.33  \\
6 & 0.25  \\
7 & 0.2  \\
8 & 0.167  \\
9 & 0.143  \\
10 & 0.125  \\
\end{tabular}
\caption{Parameter $q$ in $N$-spin RKKY helix simulations.}
\label{tab:SI_qparams}
\end{table}

Lastly, Fig.~\ref{fig:SI_helix_powerlaws}e shows the field dependence of the frequency for a 6-fold RKKY helix with biquadratic exchange, along with a fit to $f(H) = AH^3$. We take $q = 0.25, C_1 = C_2 = -   0.5$. Taken together, these results suggest that the power law behavior of the Goldstone mode gap is a model-independent feature of helices.

\bibstyle{apsrev4-1}

\end{document}